\documentclass[11pt]{article}
\usepackage[a4paper,hmargin=1.0in,vmargin=1.0in]{geometry}
\usepackage{mathtools,amsthm,amssymb,setspace,sectsty,xcolor}  
\usepackage[round]{natbib}
\usepackage[linktocpage=true,pagebackref=true,colorlinks,allcolors=blue,bookmarks=true]{hyperref}

\newtheoremstyle{DStheorem}% name of the style to be used
  {\topsep}% measure of space to leave above the theorem. E.g.: 3pt
  {\topsep}% measure of space to leave below the theorem. E.g.: 3pt
  {\itshape}% name of font to use in the body of the theorem
  {0pt}% measure of space to indent
  {\scshape}% name of head font
  {.}% punctuation between head and body
  { }% space after theorem head; " " = normal interword space
  {\thmname{#1}\thmnumber{ #2}\thmnote{ (#3)}}
\theoremstyle{DStheorem}
\newtheorem{theorem}{Theorem}[section]
\newtheorem{lemma}[theorem]{Lemma}
\newtheorem{claim}[theorem]{Claim}

\let\oldproofname=\proofname
\renewcommand{\proofname}{\rm\sc{\oldproofname}}

\newcommand{\bs}[1]{\boldsymbol{#1}}
\newcommand{\bbR}{\mathbb{R}}
\newcommand{\bbN}{\mathbb{N}}
\newcommand{\eps}{\epsilon}
\newcommand{\opt}{\mathrm{OPT}}

\newcommand{\exsub}[2]{\mathbb{E}_{#1}\left[#2\right]}

\newcommand{\prsub}[2]{\mathrm{Pr}_{#1}\left[#2\right]}

\newcommand{\poly}{\mathrm{poly}}
\newcommand{\myeoq}{\mathrm{EOQ}}
\newcommand{\mydense}{\mathrm{dense}}
\newcommand{\mysparse}{\mathrm{sparse}}
\newcommand{\myzero}{\mathrm{zero}}
\newcommand{\myallow}{\mathrm{allow}}

\sectionfont{\large} \subsectionfont{\normalsize}
\allowdisplaybreaks
\begin{document}

\begin{titlepage}

\title{Near-Optimal Dynamic Policies for \\ Joint Replenishment in Continuous/Discrete Time}
\author{%
Danny Segev\thanks{School of Mathematical Sciences and Coller School of Management, Tel Aviv University, Tel Aviv 69978, Israel. Email: {\tt segevdanny@tauex.tau.ac.il}. Supported by Israel Science Foundation grant 1407/20.}}
\date{}
\maketitle

\pagenumbering{Roman}

\begin{abstract}
While dynamic policies have historically formed the foundation of most influential papers dedicated to the  joint replenishment problem, we are still facing profound gaps in our structural understanding of optimal such policies as well as in their surrounding computational questions. To date, the seminal work of \citet{Roundy85, Roundy86}, \citet{JacksonMM85}, \citet{MaxwellM85}, and \citet{MuckstadtR87} remains unsurpassed in efficiently developing provably-good dynamic policies in this context. 

The principal contribution of this paper consists in developing a wide range of algorithmic ideas and analytical insights around the continuous-time joint replenishment problem, culminating in a deterministic framework for efficiently approximating optimal dynamic policies to any desired level of accuracy. These advances enable us to derive a compactly-encoded replenishment policy whose long-run average cost is within
factor $1 + \eps$ of the dynamic optimum, arriving at an efficient polynomial-time approximation scheme (EPTAS). Technically speaking, our approach hinges on affirmative resolutions to two fundamental open questions:
\begin{itemize}
    \item {\em Feasibility of scalable discretization}: We devise the first efficient discretization-based framework for approximating the joint replenishment problem. Specifically, we prove that every continuous-time infinite-horizon instance can be reduced to a corresponding discrete-time $O( \frac{ n^3 }{ \eps^6 } )$-period instance, while incurring a multiplicative optimality loss of at most $1 + \eps$. 

    \item {\em Enhanced guarantees for the discrete setting}: Motivated by this relation, we substantially improve on the $O( 2^{2^{O(1/\eps)}} \cdot (nT)^{ O(1) } )$-time approximation scheme of \cite{NonnerS13} for the discrete-time joint replenishment problem. Beyond an exponential improvement in running time, we demonstrate that the key pillars of their methodology --- randomization and hierarchical decompositions --- can be entirely avoided, while concurrently offering a relatively simple analysis. 
\end{itemize}

\end{abstract}


\end{titlepage}

% CONTENTS %%%%%%%%%%%%%%%%%%%%%%%%%%%%%%%%%%%%%%%
\setcounter{page}{2}
\tableofcontents

% SECTIONS %%%%%%%%%%%%%%%%%%%%%%%%%%%%%%%%%%%%%%%%%%%%%%%%%%%%%%%%%%%
\newpage
\pagestyle{plain}
\pagenumbering{arabic}
\setcounter{page}{1}
\section{Introduction}

With origins tracing back to the mid-1960s, and with stylized prototypes appearing even earlier, the joint replenishment problem has served as a foundational construct in the development of inventory theory, and concurrently, as an ideal springboard  for the birth of extremely versatile algorithmic ideas. Across the last six decades, such environments have proven themselves to be indispensable analytical frameworks, not only for advancing theoretical insights but also for reinforcing the practical applicability of inventory management, particularly in optimizing distribution centers, retail supply chains, production lines, and automotive assembly facilities. This enduring interest is reflected in an expansive body of literature that spans provably-good algorithmic  approaches, heuristic procedures, empirical investigations, real-world implementations, and decision-support software systems. As such, beyond discussing directly-related results, comprehensively reviewing even a minuscule fraction of this literature would be far exceeding the scope of our work. For a more thorough understanding of the joint replenishment landscape, we refer interested readers to well-known survey articles \citep{AksoyE88, GoyalS89, MuckstadtR93, KhoujaG08, BastosMNMC17} and book chapters 
\citep{SilverP85, Zipkin00, MuckstadtS10, SimchiLeviCB14}, as well as to the references therein.

While one could discover a fair amount of variability in their operational principles, all models studied under the joint replenishment umbrella are fundamentally concerned with the lot sizing of multiple items along a given planning horizon. Moreover, the overarching objective in these settings is to identify a replenishment policy that minimizes long-run average (or cumulative) operating costs, which typically include item-specific ordering and inventory holding costs, in parallel to joint ordering costs, incurred whenever an order is placed, regardless of its precise composition. Even though this objective may appear somewhat uncomplicated to the untrained eye, its convoluted interplay between item-specific decisions and joint ordering costs introduces distinct algorithmic challenges and poses long-standing analytical questions regarding the intrinsic structure of optimal policies and their computational tractability. At the heart of these challenges lies the necessity to efficiently synchronize multiple economic order quantity (EOQ) models, complicated by our age-old inability to come up with lossless transformations, decompositions, or reductions to simpler subproblems, in light of the inherent dependencies across different items.

\subsection{Model formulation} \label{subsec:model_definition}

In order to rigorously articulate the central research questions addressed in this paper and to lay a solid foundation for the exposition of our principal contributions, we begin by presenting a formal mathematical formulation of the joint replenishment problem in its most general continuous-time setting as well as in its discrete-time analog. While such formulations can be dichotomized along multiple dimensions, we will intentionally take the perspective of dynamic-versus-periodic replenishment, which will be particularly instructive for our contextual purposes. 

\paragraph{The economic order quantity model.} To describe the inner-workings of continuous-time joint replenishment in an accessible way, we start off by  introducing its basic building block, the economic order quantity (EOQ) model. At a high level, the latter setting seeks to identify an optimal replenishment policy for a single item, aiming to minimize its long-run average cost over the continuous planning horizon $[0,\infty)$. This item is characterized by a stationary  demand rate; as explained later on, without loss of generality, we can assume a consumption rate of one unit per unit of time. Our potentially-dynamic policy is required to completely meet this demand upon occurrence, meaning that lost sales or back orders are not permitted. In this context, the notion of a ``dynamic'' policy ${\cal P}$ will be captured by two features: 
\begin{itemize}
    \item An sequence of ordering points $0 = \tau_0 < \tau_1 < \tau_2 < \cdots$ covering the entire planning horizon $[0,\infty)$, in the sense that $\lim_{k \to \infty} \tau_k = \infty$.

    \item The real-valued ordering quantities $q_0, q_1, q_2, \ldots$ at these points.
\end{itemize}
For ease of notation, $I( {\cal P}, t )$ will stand for the inventory level we end up with at time $t$, upon implementing the replenishment policy ${\cal P}$. One straightforward way to express this function is via the relation $I( {\cal P}, t ) = \sum_{k \geq 0: \tau_k \leq t} q_k - t$. To parse this expression, note that $\sum_{k \geq 0: \tau_k \leq t} q_k$ represents the total ordering quantity up to and including time $t$; this term consists of finitely-many summands, since $\lim_{k \to \infty} \tau_k = \infty$. The second term, $t$, is precisely the cumulative consumption up until this time, due to having a unit demand rate. As such, we are  meeting our demand at any point in time via previously-placed orders if and only if $I( {\cal P}, t ) \geq 0$ for all $t \in [0,\infty)$. Any policy satisfying this condition is said to be feasible, and we designate the collection of all such policies by ${\cal F}$.

We proceed by explaining how the long-run average cost of any policy ${\cal P} \in {\cal F}$ is structured. For this purpose, each of the ordering points $\tau_0, \tau_1 ,\tau_2, \ldots$ incurs a fixed set-up cost of $K$, regardless of its quantity. In other words, letting $N( {\cal P}, [0,t)) = | \{ k \in \bbN_0 : \tau_k < t \} |$ be the number of orders placed along $[0,t)$, this interval is associated with a total ordering cost of ${\cal K}( {\cal P}, [0,t)) = K \cdot N( {\cal P}, [0,t))$. Concurrently, we are facing a linear holding cost of $2H$, incurred per time unit for each inventory unit in stock; this definition explains why we can assume a unit demand rate, simply by scaling $H$. Technically speaking, ${\cal H}( {\cal P}, [0,t))$ will denote our total holding cost across the interval $[0,t)$, given by ${\cal H}( {\cal P}, [0,t)) = 2H \cdot \int_{[0,t)} I( {\cal P}, \tau ) \mathrm{d} \tau$. Putting both ingredients into a single objective, the combined cost of a feasible policy ${\cal P}$ along $[0,t)$ will be designated by $C( {\cal P}, [0,t)) = {\cal K}( {\cal P}, [0,t)) + {\cal H}( {\cal P}, [0,t))$. In turn, its long-run average cost can be expressed as 
\begin{equation} \label{eqn:long_run_cost_single}
C( {\cal P} ) ~~=~~ \limsup_{t \to \infty} \frac{ C( {\cal P}, [0,t)) }{ t } \ .    
\end{equation}
As a side note, one can easily come up with feasible policies ${\cal P}$ for which $\lim_{t \to \infty} \frac{ C( {\cal P}, [0,t)) }{ t }$ does not exist, meaning that $\limsup$ is indeed required.

Based on the preceding description, in the economic order quantity problem, our goal is to identify a feasible replenishment policy ${\cal P}^*$ whose long-run average cost is minimized, in the sense that $C( {\cal P}^* ) = \min_{ {\cal P} \in {\cal F}} C( {\cal P} )$. By consulting relevant textbooks, such as those of \citet[Sec.~3]{Zipkin00}, \citet[Sec.~2]{MuckstadtS10}, or \citet[Sec.~7.1]{SimchiLeviCB14}, unfamiliar readers will quickly find out that the latter minimum is indeed attained, via the  extremely simple class of ``stationary order sizes and stationary intervals'' policies, to which we refer as being ``periodic'' for short. These policies are characterized by a single parameter, $T$, standing for the uniform time interval between successive orders. Having fixed this parameter, orders will be placed at the time points $0, T, 2T, 3T, \ldots$, each consisting of exactly $T$ units, meaning that zero-inventory levels are reached at each of these points. As such, noting that the long-run average cost of a periodic policy $T$ can easily be simplified to $C(T) = \frac{ K }{ T } + HT$, the economic order quantity problem admits an elegant closed-form solution. The next claim summarizes several well-known properties exhibited by this function, all following from elementary
calculus arguments.

\begin{claim} \label{clm:EOQ_properties}
Over $(0,\infty)$, the cost function $C(T) = \frac{ K }{ T } + HT$ is strictly convex. Its  unique minimizer is $T^* = \sqrt{ K / H }$, attaining an optimal cost of $C( T^* ) = 2 \sqrt{KH}$.
\end{claim}

\paragraph{The joint replenishment problem.} Given the optimality of periodic policies for the economic order quantity model, one can retrospectively wonder whether placing dynamic policies at the heart of our exposition makes any sense. Circling back to this issue in the sequel, we proceed by explaining how these foundations allow for a smooth transition to the joint replenishment problem, whose essence can be encapsulated into the next high-level question:
\begin{quote}
{\em How should we synchronize numerous economic order quantity models, when different items are interacting via joint ordering costs?}
\end{quote}
Specifically, we wish to determine and synchronize the replenishment processes of $n$ distinct items, where each item $i \in [n]$ is coupled with its own EOQ model, parameterized by ordering
and holding costs $K_i$ and $2H_i$, respectively. Along the lines of  representation~\eqref{eqn:long_run_cost_single}, deciding to replenish this item via the dynamic policy ${\cal P}_i \in {\cal F}$ would lead to a marginal long-run operating cost of $C_i( {\cal P}_i ) = \limsup_{t \to \infty} \frac{ C_i( {\cal P}_i, [0,t)) }{ t }$. However, as alluded to above, further complexity comes in the form of a joint ordering cost, $K_0$, incurred whenever an order is placed, regardless
of its particular subset of items.

Given these ingredients, it is convenient to represent any joint replenishment policy through a vector ${\cal P} = ({\cal P}_1, \ldots, {\cal P}_n)$, where ${\cal P}_i$ stands for the dynamic policy we employ with respect to item $i \in [n]$. For such policies, the first component of our objective function encloses  the sum of marginal EOQ-based costs, $\sum_{i \in [n]} C_i( {\cal P}_i )$. The second component, which will be designated by $J({\cal P})$, captures long-run average joint ordering costs. Formally, this term is defined as 
\begin{equation} \label{eqn:joint_cost_def}
J( {\cal P} ) ~~=~~ K_0 \cdot \limsup_{t \to \infty} \frac{ N({\cal P},[0,t)) }{ t } \ ,
\end{equation}
where $N({\cal P},[0,t))$ stands for the number of joint orders occurring across $[0,t)$ with respect to the policy ${\cal P}$. In summary, our goal is to determine a joint replenishment policy ${\cal P} = ({\cal P}_1, \ldots, {\cal P}_n)$ that minimizes long-run average operating costs, given by
\[ F({\cal P}) ~~=~~ J({\cal P}) + \sum_{i \in [n]} C_i( {\cal P}_i ) \ . \]
Once again, it is worth mentioning  that $\min_{{\cal P} \in {\cal F}^n} F( {\cal P} )$ is indeed attained by some feasible policy. To verify this claim, which has been an open question for several decades, avid readers may wish to consult the work of \cite{Sun04} and \cite{AdelmanK05}.

\paragraph{The discrete setting.} As previously mentioned, a central contribution of this paper resides in forming a computationally-efficient bridge, taking us from  the continuous-time formulation described above to its discrete-time counterpart. While there are multiple ways to introduce the latter setting, it is convenient to think of $[0,T)$ as our planning horizon, during which order placements are limited to integer-valued time points. In other words, $[0,T)$ is partitioned into $T$ periods, with each period $t \in [T]$ representing the segment $[t-1,t)$; henceforth, placing an order in this period would mean that it actually occurs at time $t-1$. 

In light of these restrictions, given any replenishment policy ${\cal P} = ({\cal P}_1, \ldots, {\cal P}_n)$, its combined joint ordering cost can be written as $J( {\cal P} ) = K_0 \cdot N( {\cal P} )$, with the convention that $N( {\cal P} )$ specifies the number of joint orders occurring across $[0,T)$. That said, unlike the continuous-time scenario, even though the marginal operational  cost $C_i( {\cal P}_i )$ of each item $i \in [n]$ still decomposes into its ordering and holding costs, ${\cal K}_i( {\cal P}_i )$ and ${\cal H}_i( {\cal P}_i )$, we aggregates these measures along the periods $1, \ldots, T$, rather than time-averaging over $[0,\infty)$. Specifically, our total ordering cost takes the form ${\cal K}_i( {\cal P}_i ) = K_i \cdot N( {\cal P}_i )$, where $N( {\cal P}_i )$ denotes the number of $i$-orders across $[0,T)$ with respect to the policy ${\cal P}_i$. As far as holding costs are concerned, suppose that $\Delta_{i,1}, \ldots, \Delta_{i,N( {\cal P}_i )}$ are the durations of these orders, meaning that  $\Delta_{i,1}$ time units elapse between the first and second $i$-orders, $\Delta_{i,2}$ units elapse between the second and third, so on and so forth. Then, the total holding cost we incur can be expressed as
\begin{equation} \label{eqn:holding_sum_square}
{\cal H}_i( {\cal P}_i ) ~~=~~ 2H_i \cdot \int_{[0,T)} I( {\cal P}_i, \tau ) \mathrm{d} \tau ~~=~~ H_i \cdot \sum_{\nu \in [N( {\cal P}_i )]} \Delta_{i,\nu}^2 \ .    
\end{equation}
As before, our objective is to identify a joint replenishment policy ${\cal P}$ of minimal total cost, given by $F({\cal P}) = J({\cal P}) + \sum_{i \in [n]} C_i( {\cal P}_i )$.

\subsection{Known results and open questions} \label{subsec:existing_work}

Perhaps unsurprisingly, even though dynamic policies have historically formed the foundation of most influential papers dedicated to continuous-time joint replenishment, we are still facing 
profound gaps in our structural understanding of optimal dynamic policies as well as in their surrounding computational questions. For these reasons, since the very inception of this problem, with persistent recognition spanning decades, dynamic policies have been excluded from direct optimization-based consideration, apart from two algorithmic approaches, which we proceed to elaborate on in the remainder of this section.

\paragraph{Approach 1: Power-of-$\bs{2}$ policies.} To date, the seminal work of \citet{Roundy85, Roundy86}, \citet{JacksonMM85}, \citet{MaxwellM85}, and \citet{MuckstadtR87} remains unsurpassed in efficiently constructing provably-good dynamic policies for the joint replenishment problem in its continuous-time formulation. Conceptually, these papers introduced ingenious techniques to leverage natural convex relaxations, whose optimal solutions are rounded to power-of-$2$ policies. Such  policies establish a common base, denoted $T_{\min}$, setting the time interval $T_i$ between successive $i$-orders to take the form $2^{ q_i } \cdot T_{\min}$, for some integer $q_i \geq 0$. Remarkably, while one may mistakenly infer that our benchmark is the periodic optimum, this synchronization strategy enables the joint optimization of $T_{\min}$ and $\{q_i\}_{i \in [n]}$, resulting in a power-of-$2$ policy whose long-run average cost is within factor $\frac{ 1 }{ \sqrt{2} \ln 2 } \approx 1.02$ of the dynamic optimum! Since their publication, these findings have come to be regarded as some of the most prominent breakthroughs in inventory management, owing to their broad applicability in both theoretical and practical contexts. As a supplementary remark, we point readers to the work of \cite{TeoB01}, which presents notably elegant methods for deriving these results through randomized rounding. 

With the above-mentioned approximation guarantee being state-of-the-art for nearly four decades even within the class of periodic policies, our recent work \citep{Segev25JRPa, Segev24JRPb} has finally broken new ground, revealing that the periodic setting can be efficiently approximated to any degree of accuracy. From a methodological standpoint, these two papers proposed a new algorithmic approach, termed $\Psi$-pairwise alignment, allowing us to identify a periodic replenishment policy whose long-run average cost is within factor $1 + \eps$ of optimal. For any $\eps > 0$, the running time of our algorithm is $O( 2^{ \tilde{O}(1/\eps) } \cdot n^{ O(1) } )$, corresponding to the notion of an efficient polynomial-time approximation scheme (EPTAS). Unfortunately, this approach crucially relies on positioning the periodic optimum as a comparative benchmark, and we are not aware of any method by which it can be harnessed to compete against the dynamic optimum.

\paragraph{Approach 2: Cyclic policies and discretization.} Yet another route for identifying near-optimal dynamic policies goes through a discretization-based framework. From this perspective, standard arguments can be employed to prove that such policies can be approximated within any desired degree of precision by repeatedly duplicating a cyclic policy, defined across a bounded-length segment (see, e.g., \citet[Sec.~5]{AdelmanK05}). Subsequently, the latter segment is discretized via a sufficiently-fine evenly-spaced grid, yielding an instance of the discrete-time joint replenishment problem. These connections served as our original motivation for exploring this model \citep{Segev14}, which was shown to be approximable within factor $1 + \eps$ by means of dimension reduction and dynamic programming, leading to an $O( (nT)^{ \tilde{O}( (\log \log T) / \eps ) } )$ time algorithm. Later on, \cite{NonnerS13} succeeded in eliminating the exponential dependence on the number of time periods, $T$, attaining a $(1+\eps)$-approximation in $O( 2^{2^{O(1/\eps)}} \cdot (nT)^{ O(1) } )$ time. Their approach integrates distinct dynamic programming techniques with randomized hierarchical decompositions, exhibiting certain technical parallels to the pioneering work of \cite{Arora98} on Euclidean network design. 

The preceding discussion may give the impression that a polynomial-time approximation scheme (PTAS) for dynamic continuous-time joint replenishment is readily available. However, upon closer examination, this assertion cannot be further from the truth. At present, when allowing for an $\eps$-bounded loss in optimality, it is still unresolved whether cyclic policies can be reduced to the discrete-time setting, such that the resulting planning horizon consists of polynomially-many periods. In particular, to our knowledge, all existing reductions give rise to $T = (n K_0 K_{\max} H_{\max} )^{ \Omega(1) }$ periods, where $K_{\max}$ and $H_{\max}$ represent the maximum ordering and holding costs across all items; regrettably, this quantity is not polynomial in the input size. 

\paragraph{Hardness results.} Although our work is primarily algorithmic in nature, it is instructive to briefly highlight known intractability results around  joint replenishment, mainly to provide a more complete picture of this landscape. Along these lines, \citet{SchulzT11} were the first to rigorously examine the computational feasibility of efficiently obtaining optimal periodic policies in continuous time. Specifically, they established that, in the so-called fixed-base setting, the existence of a polynomial-time algorithm for the joint replenishment problem would translate to an analogous result for integer factorization, thus uncovering intricate relationships with core challenges in number theory. Subsequently, \citet{CohenHillelY18} attained classical complexity-theoretic results, proving that the fixed-base setting is, in fact, strongly NP-hard. This finding was substantially simplified by \citet{SchulzT22}, who demonstrated that NP-hardness persists even in the presence of only two items. Additionally, the authors have migrated their original insight to the variable-base setting, unveiling  its polynomial-time reducibility to integer factorization. The latter result was further amplified to a strong NP-hardness proof by \citet{TuisovY20}. To the best of our knowledge, no lower bounds on the time complexity of computing optimal dynamic  policies have been proposed to date. In essence, all sources of intractability identified in the existing literature are heavily contingent upon periodicity, and moreover, we are still very much in the dark even regarding fundamental questions such as the plausibility of representing optimal dynamic policies in polynomial space. 

\paragraph{Open questions.} This state of affairs, marked by persistent roadblocks to the efficient computation of truly near-optimal dynamic policies for continuous-time joint replenishment, elicits numerous motivating questions that guide our present work. These questions, presented in a multitude of papers, textbooks, and course materials, can be concisely outlined as follows.
\begin{enumerate}
    \item \label{item:open_question_1} {\em Computational characterization}: Is the joint replenishment problem rendered APX-hard by dynamic policies? Alternatively, does this setting allow for a polynomial-time approximation scheme, similar to what has very recently been discovered for its restriction to periodic policies \citep{Segev25JRPa, Segev24JRPb}?

    \item \label{item:open_question_2} {\em Feasibility of scalable discretization}: At the expense of an $\eps$-bounded loss in optimality, is it possible to efficiently translate any $n$-item continuous-time instance into a compact discrete-time proxy, whose underlying number of periods $T$ is polynomial in $n$, $1/\eps$, and nothing beyond? 

    \item \label{item:open_question_3} {\em Enhanced guarantees for the discrete setting:} With a touch of optimism regarding question~\ref{item:open_question_2}, what prospects are there for surpassing the double-exponential $O( 2^{2^{O(1/\eps)}} \cdot (nT)^{ O(1) } )$-time approximation scheme of \cite{NonnerS13} for general instances of the discrete formulation? Moreover, could one come up with a simpler  approach, potentially sidestepping the need for randomization, hierarchical decompositions, and intricate analysis?
\end{enumerate}

\subsection{Main contributions} \label{subsec:contributions}

The principal contribution of this paper consists in developing a wide range of algorithmic ideas and analytical insights around the continuous-time joint replenishment problem, culminating in a deterministic framework for efficiently approximating optimal dynamic policies to any desired level of accuracy. As formally stated in Theorem~\ref{thm:EPTAS_dynamic_cont} below, these advances enable us to determine a compactly-encoded replenishment policy whose long-run average cost is within
factor $1 + \eps$ of the dynamic optimum, arriving at an efficient polynomial-time approximation scheme (EPTAS), wherein $\eps$-related terms are decoupled from those related to the input size. Circling back to open question~\ref{item:open_question_1}, whether examining the full spectrum of dynamic policies or the narrower class of periodic ones, we now realize that both settings admit approximation schemes of this nature, derived via surprisingly different mechanisms.

\begin{theorem} \label{thm:EPTAS_dynamic_cont}
For any $\eps > 0$, the continuous-time joint replenishment problem can be approximated within factor $1 + \eps$ of optimal. The running time of our algorithm is $O( 2^{ \tilde{O}( 1 / \eps^4 ) } \cdot |{\cal I}|^{ O(1) } )$, where $|{\cal I}|$ designates its input size in binary representation.
\end{theorem}

From a technical standpoint, the design of our approximation scheme hinges on affirmative resolutions to open questions~\ref{item:open_question_2} and~\ref{item:open_question_3}. In what follows, to shed light on how various building blocks interact, we provide a succinct account of these findings, postponing detailed discussions of their structural, algorithmic, and analytical aspects to subsequent sections. 

\paragraph{First milestone: Feasibility of scalable discretization.} In Section~\ref{sec:reduction}, we devise an efficient discretization-based framework for approximating the joint replenishment problem, proving that every continuous-time infinite-horizon instance can be reduced to the discrete-time finite-horizon setting, while incurring a multiplicative optimality loss of at most $1 + \eps$. In pursuit of this objective, we present a sequence of structural transformations, meant to instill important ``regularity'' properties into free-form optimal dynamic policies. As a byproduct, our resulting discrete-time instance is constructed by subdividing a carefully chosen closed segment via $T = O( \frac{ n^3 }{ \eps^6 } )$ evenly-spaced points, yielding the first-ever reduction to polynomially-many periods. 

\paragraph{Second milestone: Enhanced guarantees for the discrete setting.} With discrete-time finite-horizon approximations now established as ready-to-deploy pathways for computing near-optimal dynamic policies in continuous time, we turn our attention to improving upon state-of-the-art results in this context. In Section~\ref{sec:EPTAS}, we uncover new insights into group-fusing,   grid-restricted policies, and  several lower-bounding gadgets to substantially improve on the $O( 2^{2^{O(1/\eps)}} \cdot (nT)^{ O(1) } )$-time approximation scheme of \cite{NonnerS13}. Beyond an exponential improvement in running time, whose finer details are formally stated below, we demonstrate that randomization and hierarchical decompositions can be  avoided altogether, while concurrently offering a relatively simple analysis. 

\begin{theorem} \label{thm:main_EPTAS}
For any $\eps > 0$, the discrete-time joint replenishment problem can be approximated within factor $1 + \eps$ of optimal. The running time of our algorithm is $O( 2^{ \tilde{O}( 1 / \eps^4 ) } \cdot (nT)^{ O(1) } )$.    
\end{theorem}
\section{Feasibility of Scalable Discretization} \label{sec:reduction}

In what follows, we devise a polynomial-time reduction, mapping every continuous-time infinite-horizon instance of the joint replenishment problem to the discrete-time finite-horizon setting, while blowing-up its long-run average cost by a factor of at most $1 + \eps$. For this purpose, along the next few sections, we present a sequence of structural transformations, gradually creating a discrete-time instance defined across $T = O( \frac{ n^3 }{ \eps^6 } )$ uniform-length periods. This result forms a computationally-efficient bridge, taking us from the continuous-time formulation to its discrete-time counterpart, which is the topic of Section~\ref{sec:EPTAS}.

\subsection{Step 1: Polynomially-related ordering costs} \label{subsec:poly_related_Ki}

Let ${\cal P}^* = ({\cal P}^*_1, \ldots, {\cal P}^*_n)$ be an optimal dynamic policy with respect to a given instance of the continuous-time formulation. We say that item $i$ is expensive when $K_i > \frac{ K_0 }{ \eps }$; in the opposite direction, when $K_i < \frac{ \eps K_0 }{n}$, this item is called cheap. In what follows, we describe a rather straightforward reduction to instances excluding expensive and cheap items, meaning that all individual ordering costs reside within $[\frac{ \eps K_0 }{n}, \frac{ K_0 }{ \eps }]$. This step will increase our long-run average cost by a factor of at most $1 + 2\eps$.

\paragraph{Handling expensive items.} To begin, we claim that all expensive  items can be independently treated, increasing our cost by an additive factor of at most $\eps \cdot \sum_{i \in [n]} C_i( {\cal P}_i^* )$. Indeed, for any such item $i$, suppose we compute an optimal EOQ policy ${\cal P}_i^{ \myeoq }$, which is known to be periodic, with an ordering interval of $T^*_i = \sqrt{ K_i / H_i }$, as explained in Section~\ref{subsec:model_definition}. When each of the resulting $i$-orders is augmented with a separate joint ordering cost of $K_0$, our marginal long-run average cost becomes 
\[ \frac{ K_0 + K_i }{ T^*_i } + H_i T^*_i ~~<~~ (1 + \eps) \cdot \frac{ K_i }{ T^*_i } + H_i T^*_i ~~\leq~~ (1 + \eps) \cdot C_i( {\cal P}_i^{ \myeoq } ) ~~\leq~~ (1 + \eps) \cdot C_i( {\cal P}_i^* ) \ . \]
Here, the first inequality holds since $K_i > \frac{ K_0 }{ \eps }$, as item $i$ is expensive, and the third inequality follows by recalling that ${\cal P}_i^{ \myeoq }$ is a minimizer of the long-run cost function $C_i( \cdot )$ across all dynamic policies.

\paragraph{Handling cheap items.} Next, knowing that the ordering cost $K_i$ of every cheap item $i$ resides in $[0,\frac{ \eps K_0 }{n})$, suppose we increment this parameter to $\hat{K}_i = \frac{ \eps K_0 }{n}$. As a result, with respect to its component ${\cal P}_i^*$ in the optimal policy ${\cal P}^*$, the marginal long-run average cost of this item turns into
\begin{eqnarray*}
\hat{C}_i( {\cal P}_i^* ) & = & \lim_{t \to \infty} \frac{ \hat{K}_i \cdot N( {\cal P}_i^*, [0,t)) + {\cal H}( {\cal P}_i^*, [0,t)) }{ t } \\
& \leq & \frac{ \eps K_0 }{ n } \cdot \lim_{t \to \infty} \frac{ N( {\cal P}^*, [0,t)) }{ t } + \lim_{t \to \infty} \frac{ {\cal H}( {\cal P}_i^*, [0,t)) }{ t } \\
& \leq & \frac{ \eps }{ n } \cdot J( {\cal P}^* ) + C_i( {\cal P}_i^* ) \ .
\end{eqnarray*}
Summing over all cheap items, it follows that this modification increases our overall long-run average cost by at most $\eps \cdot J( {\cal P}^* ) \leq \eps \cdot F( {\cal P}^* )$. For simplicity of presentation, due to ending up with a modified instance as a consequence of the current reduction, we recycle ${\cal P}^*$ for the purpose of referring to an optimal policy in this context. 

\subsection{Step 2: Near-optimal cyclic policy}

As mentioned in Section~\ref{subsec:existing_work}, for any $\eps > 0$, the continuous-time joint replenishment problem has long been known to admit $\eps$-optimal cyclic policies; see, e.g., \cite{AdelmanK05}. However, for algorithmic reasons that will be clarified later on, this finding by itself would not suffice. In particular, we wish to establish the existence of an $\eps$-optimal cyclic policy whose planning horizon length is appropriately bounded in terms of several input parameters, which are compacted into ${\cal M} = K_0 + n \cdot (K_{\max} + H_{\max})$. The precise guarantees we attain along these lines are stated in the next claim, whose proof is presented in the remainder of this section.  

\begin{lemma} \label{lem:reduction_cyclic_eps}
There exists a segment ${\cal S}$ of length $\| {\cal S} \| \in [\frac{ K_{\max} }{ 4\eps {\cal M}}, \frac{ 16 {\cal M} }{ \eps^2 H_{\min} }]$, where an optimal policy ${\cal P}^{{\cal S}*}$ has a time-average cost of $\frac{ F( {\cal P}^{{\cal S}*}, {\cal S} ) }{ \| {\cal S} \| } \leq (1 + \eps) \cdot F( {\cal P}^* )$.
\end{lemma}

\paragraph{Identifying the segment $\bs{\cal S}$.} With respect to the optimal policy ${\cal P}^*$, let us define an infinite sequence $\tau_1 < \tau_2 < \cdots$ of time points as follows:
\begin{itemize}
    \item First, $\tau_1$ is the minimal time point $t > 0$ for which the segment $[0,t]$ has at least $\frac{ 1 }{ \eps } + 1$ orders of every item $i \in [n]$, along with exactly $\frac{ 1 }{ \eps } + 1$ orders for some item $i \in [n]$.

    \item Next, $\tau_2$ is the minimal $t > \tau_1$ for which the segment $[\tau_1,t]$ meets this condition.

    \item The remaining sequence $\tau_3, \tau_4, \ldots$ is defined in a similar way.
\end{itemize}
By recalling how dynamic replenishment policies are defined (see Section~\ref{subsec:model_definition}), it is easy to verify that the sequence $\tau_1, \tau_2, \ldots$ is well-defined and tends to infinity. 

Given this sequence, one way of expressing the optimal long-run average cost is via $F( {\cal P^*} ) = \lim_{\hat{\kappa} \to \infty} \frac{ F( {\cal P}^*, [0, \tau_{\hat{\kappa}} )) }{ \tau_{\hat{\kappa}} }$, noting that 
$\tau_{\hat{\kappa}} = \sum_{ \kappa \leq \hat{\kappa} } ( \tau_{\kappa} - \tau_{\kappa-1} )$ and $F( {\cal P}^*, [0, \tau_{\hat{\kappa}} )) = \sum_{ \kappa \leq \hat{\kappa} } F( {\cal P}^*, [\tau_{\kappa-1}, \tau_{\kappa} ))$. Consequently, for any $\eps >0$, there exists an index $\kappa_{\eps}$ for which 
\begin{equation} \label{eqn:proof_lem_good_cyclic_eq1}
\frac{ F( {\cal P^*}, [ \tau_{\kappa_{\eps}}, \tau_{\kappa_{\eps}+1}) ) }{ \tau_{\kappa_{\eps}+1} - \tau_{\kappa_{\eps}} } ~~\leq~~ (1 + \eps) \cdot F( {\cal P^*} ) \ ,
\end{equation}
and our desired segment will be picked as ${\cal S}= [ \tau_{\kappa_{\eps}}, \tau_{\kappa_{\eps}+1})$.

\paragraph{The policy $\bs{{\cal P}^{\cal S}}$.} Rather than analyzing an optimal policy for this segment, it is convenient to consider a feasible policy ${\cal P}^{\cal S}$, created by modifying the restriction of ${\cal P}^*$ to ${\cal S}$ as follows. First, we place an order of each item $i \in [n]$ at the left endpoint $\tau_{\kappa_{\eps}}$ of ${\cal S}$, consisting of $I( {\cal P}^*_i, \tau_{\kappa_{\eps}})$ units. By definition of $\tau_{\kappa_{\eps}+1}$, the optimal policy ${\cal P}^*$  places at least $\frac{ 1 }{ \eps }$ orders of each item in ${\cal S}$, meaning that this modification increases our ordering cost across ${\cal S}$ by a factor of at most $1 + \eps$. Next, we decrement the last order of each item, so that zero inventory is reached at the right endpoint $\tau_{\kappa_{\eps}+1}$ of ${\cal S}$; this operation clearly leads to a holding cost decrement. All in all, we end up with a replenishment policy ${\cal P}^{\cal S}$ whose  time-average cost is  
\begin{equation} \label{eqn:cost_bound_segment}
\frac{ F( {\cal P}^{\cal S}, {\cal S} ) }{ \| {\cal S} \| } ~~\leq~~ (1 + \eps) \cdot \frac{ F( {\cal P^*}, [ \tau_{\kappa_{\eps}}, \tau_{\kappa_{\eps}+1}) ) }{ \tau_{\kappa_{\eps}+1} - \tau_{\kappa_{\eps}} } ~~\leq~~ (1 + \eps)^2 \cdot F( {\cal P^*} ) \ ,   
\end{equation}
where the last inequality is obtained by plugging in~\eqref{eqn:proof_lem_good_cyclic_eq1}.

\paragraph{Bounding $\bs{\| {\cal S} \|}$.} To conclude the proof of Lemma~\ref{lem:reduction_cyclic_eps}, it remains to show that the segment ${\cal S}$ is of length $\| {\cal S} \| \in [\frac{ K_{\max} }{ 4\eps {\cal M}}, \frac{ 16 {\cal M} }{ \eps^2 H_{\min} }]$. For this purpose, let us recall that by definition, ${\cal M} = K_0 + n \cdot (K_{\max} + H_{\max})$. Our motivation for making this particular choice is that ${\cal M}$ constitutes a straightforward upper bound on the optimal long-run average cost, $F( {\cal P^*})$. Indeed, with respect to the infinite-horizon setting, one possible policy consists of ordering all items at all integer time points, implying that  
\begin{equation} \label{eqn:opt_cost_vs_M}
F( {\cal P^*}) ~~\leq~~ K_0 + \sum_{i \in [n]} \left( K_i + H_i \right) ~~\leq~~ K_0 + n \cdot \left( K_{\max} + H_{\max}  \right) ~~=~~ {\cal M} \ . 
\end{equation}
Given this observation, we proceed to bound the length of ${\cal S}$ as follows.
\begin{itemize}
    \item {\em Upper bound}: By definition of $\tau_{\kappa_{\eps}+1}$, we know that there is at least one item $i \in [n]$ with $N( {\cal P}^{\cal S}_i, {\cal S}) \leq \frac{ 1 }{ \eps } + 1$ orders across ${\cal S}$. Since at least one of these orders has a duration of at least $\frac{ \| {\cal S} \| }{ N( {\cal P}^{\cal S}_i, {\cal S})} \geq \frac{ \eps }{2} \cdot \| {\cal S} \|$, it incurs a holding cost of at least $H_i \cdot ( \frac{ \eps }{2} \cdot \| {\cal S} \| )^2$. As a result,
    \[ H_i \cdot \left( \frac{ \eps }{2} \cdot \| {\cal S} \| \right)^2 ~~\leq~~ F( {\cal P}^{\cal S}, {\cal S} ) ~~\leq~~ (1 + \eps)^2 \cdot \| {\cal S} \| \cdot F( {\cal P^*} ) ~~\leq~~ (1 + \eps)^2 \cdot \| {\cal S} \| \cdot {\cal M} \ , \]
    where the second and third inequalities follow from~\eqref{eqn:cost_bound_segment} and~\eqref{eqn:opt_cost_vs_M}, respectively. By rearranging, $\| {\cal S} \| \leq \frac{ 4(1 + \eps)^2 \cdot {\cal M} }{ \eps^2 H_i } \leq \frac{ 16 {\cal M} }{ \eps^2 H_{\min} }$.

    \item {\em Lower bound}: Again by definition of $\tau_{\kappa_{\eps}+1}$,  every item $i \in [n]$ has $N( {\cal P}^{\cal S}_i, {\cal S}) \geq \frac{ 1 }{ \eps }$ orders across ${\cal S}$. Therefore, ${\cal P}^{\cal S}$ pays at least $\frac{ K_{\max} }{ \eps }$ along this segment, implying that $\frac{ K_{\max} }{ \eps } \leq F( {\cal P}^{\cal S}, {\cal S} ) \leq (1 + \eps)^2 \cdot \| {\cal S} \| \cdot {\cal M}$. By rearranging, $\| {\cal S} \| \geq \frac{ K_{\max} }{ \eps(1+\eps)^2 \cdot {\cal M}} \geq \frac{ K_{\max} }{ 4\eps {\cal M}}$.
\end{itemize}

\subsection{Step 3: Order boosting}

Even though Lemma~\ref{lem:reduction_cyclic_eps} informs us that it suffices to focus on the segment ${\cal S}$, we cannot exclude the possibility that its optimal policy ${\cal P}^{{\cal S}*}$ could be highly imbalanced in terms of the number of times $\{ N( {\cal P}^{{\cal S}*}_i, {\cal S} ) \}_{i \in [n]}$ different items are ordered. Yet another important ingredient of our reduction is proving that ${\cal S}$ admits a near-optimal policy where these numbers are not exponentially far apart, as formally stated below.

\begin{lemma} \label{lem:reduction_order_equating}
There exists a policy ${\cal P}^{{\cal S}+}$ with respect to the segment ${\cal S}$ such that:
\begin{enumerate}
    \item {\em Near-optimal cost}: $F( {\cal P}^{{\cal S}+}, {\cal S} ) \leq (1 + \eps) \cdot F( {\cal P}^{{\cal S}*}, {\cal S} )$.

    \item {\em Balanced number of orders}: $\{ N( {\cal P}^{{\cal S}+}_i, {\cal S} ) \}_{i \in [n]}$ are within factor $\frac{ 2n^2 }{ \eps^3 }$ of each other.
\end{enumerate}
\end{lemma}
\begin{proof}
Let $i^*$ be an item whose number of orders $N_i^* = N( {\cal P}^{{\cal S}*}_i, {\cal S})$ with respect to the optimal policy ${\cal P}^{{\cal S}*}$ is maximal; when this maximum is attained by multiple items, $i^*$ is arbitrarily selected among them. To define the policy ${\cal P}^{{\cal S}+}$, for all items $i$ with $N_i^* \geq \frac{ \eps^3 }{ 2n^2 } \cdot N_{i^*}^*$ orders, we keep their policy unchanged, meaning that ${\cal P}^{{\cal S}+}_i = {\cal P}^{{\cal S}*}_i$. In contrast, for each item $i$ with $N_i^* < \frac{ \eps^3 }{ 2n^2 } \cdot N_{i^*}^*$, we convert ${\cal P}^{{\cal S}*}$ into a new policy ${\cal P}^{{\cal S}+}_i$, by adding $\lceil \frac{ \eps^3 }{ 2n^2 } \cdot N_{i^*}^* \rceil - N_i^*$ arbitrary orders, restricted to points where joint orders have already been placed. It is easy to verify that, with these additional orders, $i$-ordering quantities can be adjusted to end up with a holding cost of ${\cal H}_i( {\cal P}^{{\cal S}+}_i, {\cal S}) \leq {\cal H}_i( {\cal P}^{{\cal S}*}_i, {\cal S})$. As such, the only ingredient where ${\cal P}^{{\cal S}+}_i$ is more expensive than ${\cal P}^{{\cal S}*}_i$ is its $i$-ordering cost. To upper-bound this term, note that
\begin{eqnarray*}
{\cal K}_i( {\cal P}^{{\cal S}+}_i, {\cal S}) & = & K_i \cdot N( {\cal P}^{{\cal S}+}_i, {\cal S})  \\
& = & K_i \cdot \left\lceil \frac{ \eps^3 }{ 2n^2 } \cdot N_{i^*}^* \right\rceil  \\
& \leq & \frac{ n }{ \eps^2 } \cdot K_{i^*} \cdot \frac{ \eps^3 }{ n^2 } \cdot N_{i^*}^* \\
& = & \frac{ \eps }{ n } \cdot K_{i^*} \cdot N( {\cal P}^{{\cal S}*}_{i^*}, {\cal S})  \\
& = & \frac{ \eps }{ n } \cdot {\cal K}_{i^*}( {\cal P}^{{\cal S}*}_{i^*}, {\cal S}) \ , 
\end{eqnarray*}
where the sole inequality above  holds since, following Section~\ref{subsec:poly_related_Ki}, we have $K_1, \ldots, K_n \in [ \frac{ \eps K_0 }{n}, \frac{ K_0 }{ \eps } ]$. Therefore, we have indeed constructed a policy ${\cal P}^{{\cal S}+}$ with a total cost of $F( {\cal P}^{{\cal S}+}, {\cal S} ) \leq (1 + \eps) \cdot F( {\cal P}^{{\cal S}*}, {\cal S} )$, where $\{ N( {\cal P}^{{\cal S}+}_i, {\cal S} ) \}_{i \in [n]}$ all reside within $[\frac{ \eps^3 }{ 2n^2 } \cdot N_{i^*}^*, N_{i^*}^*]$.
\end{proof}

\subsection{Step 4: Polynomially-many orders}

The next-to-last structural property we wish to instill is that our segment of interest will admit a near-optimal policy with polynomially-many joint orders. While this property may be violated by ${\cal S}$ itself, the next claim shows that it can be guaranteed by a carefully-chosen subsegment.

\begin{lemma} \label{lem:cheap_subsegment}
There exists a segment ${\cal T} \subseteq {\cal S}$ of length $\| {\cal T} \| \in [\frac{ K_0 }{ 16 {\cal M} }, \frac{ 16 {\cal M} }{ \eps^2 H_{\min} }]$, along with  a policy ${\cal P}^{\cal T}$ for this segment, such that: 
\begin{enumerate}
    \item {\em Near-optimal cost}: $\frac{ F( {\cal P}^{{\cal T}}, {\cal T} ) }{ \| {\cal T} \| } \leq (1 + 15\eps) \cdot F( {\cal P}^* )$.     

    \item {\em $\poly(n, \frac{ 1 }{ \eps })$ joint orders}: $N( {\cal P}^{{\cal T}}, {\cal T} ) \leq \frac{ 12n^3 }{ \eps^5 }$.
\end{enumerate} 
\end{lemma}

\paragraph{Breaking $\bs{\cal S}$ into equal-length segments.} To derive this result, let $N^+ = N( {\cal P}^{{\cal S}+}, {\cal S} )$ be the number of joint orders placed by ${\cal P}^{{\cal S}+}$ across ${\cal S}$. When $N^+ < \frac{ 2n^3 }{ \eps^4 }$, by fixing ${\cal T} = {\cal S}$ we are clearly done, since
\begin{equation} \label{eqn:lem_cheap_subsegment_1}
\frac{ F( {\cal P}^{{\cal S}+}, {\cal S} ) }{ \| {\cal S} \| } ~~\leq~~ (1 + \eps) \cdot \frac{ F( {\cal P}^{{\cal S}*}, {\cal S} ) }{ \| {\cal S} \| } ~~\leq~~ (1 + 3\eps) \cdot F( {\cal P}^* ) \ ,    
\end{equation}
where the first and second inequalities respectively follow from
Lemmas~\ref{lem:reduction_order_equating} and~\ref{lem:reduction_cyclic_eps}. Therefore, in the remainder of this proof, we assume that $N^+ \geq \frac{ 2n^3 }{ \eps^4 }$. 

In this case, let us decompose the segment ${\cal S}$ into $M = \lceil \frac{ \eps^4 N^+ }{ 2n^3 } \rceil$ equal-length subsegments, ${\cal T}_1, \ldots, {\cal T}_M$. From a solution perspective, we modify the policy ${\cal P}^{{\cal S}+}$ by adding joint orders at the $M-1$ breakpoints between these segments; each of these orders contains all items. It is not difficult to verify that ordering quantities can be adjusted such that all items arrive at each breakpoint with zero inventory. The next claim upper-bounds the time-average cost of our resulting policy, $\hat{\cal P}^{{\cal S}}$.

\begin{claim} \label{clm:time_average_hatP_S}
$\frac{ F( \hat{\cal P}^{{\cal S}}, {\cal S} ) }{ \| {\cal S} \| } \leq (1 + 7\eps) \cdot F( {\cal P}^* )$.    
\end{claim}
\begin{proof}
We first observe that, due to augmenting ${\cal P}^{{\cal S}+}$ with additional orders and subsequently adjusting ordering quantities, the holding cost of each item $i \in [n]$ may only decrease, implying that ${\cal H}_i( \hat{\cal P}^{{\cal S}}_i, {\cal S}) \leq {\cal H}_i( {\cal P}^{{\cal S}+}_i, {\cal S})$. On the other hand, $\hat{\cal P}^{\cal S}$ differs from ${\cal P}^{{\cal S}+}$ in its joint ordering cost, as well as in the marginal ordering costs of the various items. To upper-bound the former term, note that $J( \hat{\cal P}^{\cal S}, {\cal S} ) \leq (1 + \eps) \cdot J( {\cal P}^{{\cal S}+}, {\cal S} )$, since the number of orders we are currently placing is
\begin{equation} \label{eqn:clm_time_average_hatP_S_2}
N( \hat{\cal P}^{\cal S}, {\cal S} ) ~~=~~ N^+ + M-1 ~~\leq~~ N^+ + \frac{ \eps^4 N^+ }{ 2n^3 } ~~\leq~~ (1 + \eps) \cdot  N( {\cal P}^{{\cal S}+}, {\cal S} ) \ .
\end{equation}
As for the ordering cost of each item $i$, we have ${\cal K}_i( \hat{\cal P}^{{\cal S}}_i, {\cal S}) \leq (1 + \eps) \cdot {\cal K}_i( {\cal P}^{{\cal S}+}_i, {\cal S})$, since 
\begin{eqnarray*}
N( \hat{\cal P}^{{\cal S}}_i, {\cal S}) & = & N( {\cal P}^{{\cal S}+}_i, {\cal S}) + M - 1  \\
& \leq & N( {\cal P}^{{\cal S}+}_i, {\cal S}) + \frac{ \eps^4 N^+ }{ 2n^3 }  \\
& \leq & (1 + \eps) \cdot  N( {\cal P}^{{\cal S}+}_i, {\cal S}) \ . 
\end{eqnarray*}
To understand the last inequality, let $i^*$ be an item whose number of orders $N( {\cal P}^{{\cal S}+}_i, {\cal S})$ with respect to ${\cal P}^{{\cal S}+}$ is maximal. Then, this item must appear in at least $\frac{ N^+ }{ n }$ of the joint orders placed by ${\cal P}^{{\cal S}+}$, implying that
\begin{equation} \label{eqn:clm_time_average_hatP_S_1}
\frac{ N^+ }{ n } ~~\leq~~ N( {\cal P}^{{\cal S}+}_{i^*}, {\cal S}) ~~\leq~~ \frac{ 2n^2 }{ \eps^3 } \cdot N( {\cal P}^{{\cal S}+}_i, {\cal S}) \ ,  
\end{equation}
where the second inequality follows from Lemma~\ref{lem:reduction_order_equating}, stating that $\{ N( {\cal P}^{{\cal S}+}_i, {\cal S} ) \}_{i \in [n]}$ are all within factor $\frac{ 2n^2 }{ \eps^3 }$ of each other. By rearranging~\eqref{eqn:clm_time_average_hatP_S_1}, it follows that $N^+ \leq \frac{ 2n^3 }{ \eps^3 } \cdot N( {\cal P}^{{\cal S}+}_i, {\cal S})$. In summary, we have just shown that
\[ \frac{ F( \hat{\cal P}^{{\cal S}}, {\cal S} ) }{ \| {\cal S} \| } ~~\leq~~ (1 + \eps) \cdot  
\frac{ F( {\cal P}^{{\cal S}+}, {\cal S} ) }{ \| {\cal S} \| } ~~\leq~~ (1 + 7\eps) \cdot F( {\cal P}^* ) \ , \]
where the last inequality holds since $\frac{ F( {\cal P}^{{\cal S}+}, {\cal S} ) }{ \| {\cal S} \| } \leq (1 + 3\eps) \cdot F( {\cal P}^* )$, by inequality~\eqref{eqn:lem_cheap_subsegment_1}.
\end{proof}

\paragraph{Focusing on a single segment.} Since the policy $\hat{\cal P}^{{\cal S}}$ arrives to each breakpoint with zero inventory for all items, its restriction  to each segment ${\cal T}_m$ defines a valid replenishment policy, which will be referred to as $\hat{\cal P}^{m}$. For the purpose of establishing Lemma~\ref{lem:cheap_subsegment}, we claim that at least one of the segments ${\cal T}_1, \ldots, {\cal T}_M$ can play the role of our desired ${\cal T}$. In what follows, we prove this claim by employing the probabilistic method \citep{AlonS2016}, noting that an alternative proof could utilize averaging arguments.

\begin{lemma} \label{lem:rand_seg_choice}
For some $m^* \in [M]$,  the segment ${\cal T}_{m^*}$ and its corresponding policy $\hat{\cal P}^{m^*}$ satisfy $\frac{ F( \hat{\cal P}^{m^*}, {\cal T}_{m^*} ) }{ \| {\cal T}_{m^*} \| } \leq (1 + 15\eps) \cdot F( {\cal P}^* )$ and $N( \hat{\cal P}^{m^*}, {\cal T}_{m^*} ) \leq \frac{ 12n^3 }{ \eps^5 }$. 
\end{lemma}
\begin{proof}
Suppose that $\mu$ is drawn uniformly at random out of $1, \ldots, M$. Then, the expected number of joint orders we are seeing across the segment ${\cal T}_{\mu}$ with respect to $\hat{\cal P}^{\mu}$ is
\begin{eqnarray}
 \exsub{ \mu }{ N( \hat{\cal P}^{\mu}, {\cal T}_{\mu} ) } & = & \frac{ N( \hat{\cal P}^{{\cal S}}, {\cal S} ) }{ M } \nonumber \\
 & \leq & (1 + \eps) \cdot \frac{ N( {\cal P}^{{\cal S}+}, {\cal S} ) }{ M } \label{eqn:lem_rand_seg_choice_1} \\
 & \leq & \frac{ 2N^+ }{ M } \nonumber \\
 & \leq & \frac{ 4n^3 }{ \eps^4 } \ . \label{eqn:lem_rand_seg_choice_2} 
\end{eqnarray}
Here, inequality~\eqref{eqn:lem_rand_seg_choice_1} is obtained by plugging in~\eqref{eqn:clm_time_average_hatP_S_2}, whereas inequality~\eqref{eqn:lem_rand_seg_choice_2} holds since $M = \lceil \frac{ \eps^4 N^+ }{ 2n^3 } \rceil$. In terms of expected time-average cost, we have
\begin{equation} \label{eqn:lem_rand_seg_choice_3}
\exsub{ \mu }{ \frac{ F( \hat{\cal P}^{\mu}, {\cal T}_{\mu} ) }{ \| {\cal T}_{\mu } \| } }  ~~=~~\frac{ F( \hat{\cal P}^{{\cal S}}, {\cal S} ) }{ \| {\cal S} \| } ~~\leq ~~ (1 + 7\eps) \cdot F( {\cal P}^* ) \ , 
\end{equation}
where the last inequality is precisely the result stated in Claim~\ref{clm:time_average_hatP_S}. Therefore,
\begin{eqnarray}
&& \prsub{ \mu }{ \left\{ \frac{ F( \hat{\cal P}^{\mu}, {\cal T}_{\mu} ) }{ \| {\cal T}_{\mu } \| } \leq (1 + 15\eps) \cdot F( {\cal P}^* ) \right\} \wedge \left\{ N( \hat{\cal P}^{\mu}, {\cal T}_{\mu} ) \leq \frac{ 12n^3 }{ \eps^5 } \right\} } \nonumber \\
&& \qquad \geq~~ 1 - \prsub{ \mu }{ \frac{ F( \hat{\cal P}^{\mu}, {\cal T}_{\mu} ) }{ \| {\cal T}_{\mu } \| } > (1 + 15\eps) \cdot F( {\cal P}^* ) } - \prsub{ \mu }{ N( \hat{\cal P}^{\mu}, {\cal T}_{\mu} ) > \frac{ 12n^3 }{ \eps^5 } } \nonumber \\
&& \qquad \geq~~ 1 - \prsub{ \mu }{ \frac{ F( \hat{\cal P}^{\mu}, {\cal T}_{\mu} ) }{ \| {\cal T}_{\mu } \| } > (1 + \eps) \cdot \exsub{ \mu }{ \frac{ F( \hat{\cal P}^{\mu}, {\cal T}_{\mu} ) }{ \| {\cal T}_{\mu } \| } } } \nonumber \\
&& \qquad \qquad \qquad \qquad \qquad \mbox{} - \prsub{ \mu }{ N( \hat{\cal P}^{\mu}, {\cal T}_{\mu} ) > \frac{ 3 }{ \eps } \cdot \exsub{ \mu }{ N( \hat{\cal P}^{\mu}, {\cal T}_{\mu} ) } } \label{eqn:lem_rand_seg_choice_4}  \\
&& \qquad \geq~~ 1 - \frac{ 1 }{ 1 + \eps } - \frac{ \eps }{ 3 } \label{eqn:lem_rand_seg_choice_5} \\
&& \qquad >~~ 0 \ . \nonumber
\end{eqnarray}
Here, inequality~\eqref{eqn:lem_rand_seg_choice_4} follows from~\eqref{eqn:lem_rand_seg_choice_2} and~\eqref{eqn:lem_rand_seg_choice_3}, whereas inequality~\eqref{eqn:lem_rand_seg_choice_5} is attained via Markov's inequality. Therefore, there exists some $m^* \in [M]$ for which $\frac{ F( \hat{\cal P}^{m^*}, {\cal T}_{m^*} ) }{ \| {\cal T}_{m^*} \| } \leq (1 + 15\eps) \cdot F( {\cal P}^* )$ and $N( \hat{\cal P}^{m^*}, {\cal T}_{m^*} ) \leq \frac{ 12n^3 }{ \eps^5 }$.
\end{proof}

\paragraph{Length bounds.} Since ${\cal T}_{m^*} \subseteq {\cal S}$, we have $\| {\cal T}_{m^*} \| \leq \| {\cal S} \| \leq \frac{ 16 {\cal M} }{ \eps^2 H_{\min} }$, by Lemma~\ref{lem:reduction_cyclic_eps}. To obtain a lower bound on $\| {\cal T}_{m^*} \|$, note that $F( \hat{\cal P}^{m^*}, {\cal T}_{m^*} ) \geq K_0$, as $\hat{\cal P}^{m^*}$ is required to place at least one joint order in ${\cal T}_{m^*}$. Therefore, 
\[ \frac{ K_0 }{ \| {\cal T}_{m^*} \| } ~~\leq~~ \frac{ F( \hat{\cal P}^{m^*}, {\cal T}_{m^*} ) }{ \| {\cal T}_{m^*} \| } ~~\leq~~ (1 + 15\eps) \cdot F( {\cal P}^* ) ~~\leq~~ 16 {\cal M} \ , \]
where the second inequality corresponds to the first property stated in Lemma~\ref{lem:rand_seg_choice}, and the third inequality holds since $F( {\cal P}^* ) \leq {\cal M}$, as shown in~\eqref{eqn:opt_cost_vs_M}. By rearranging this bound, $\| {\cal T}_{m^*} \| \geq \frac{ K_0 }{ 16 {\cal M} }$.

\subsection{Step 5: Discretization} \label{subsec:reduc_discretization}

With Lemma~\ref{lem:cheap_subsegment} in place, we are ready to put the last piece of the puzzle, showing that the continuous-time joint replenishment problem with respect to ${\cal T}$ can be $\eps$-approximated by a discrete-time instance, created by an evenly-space grid in this segment. 

\paragraph{Grid-restricted policies.} Specifically, by Lemma~\ref{lem:cheap_subsegment}, we know that the policy ${\cal P}^{\cal T}$ places $N^{\cal T} = N( {\cal P}^{\cal T}, {\cal T} ) \leq \frac{ 12n^3 }{ \eps^5 }$ joint orders across ${\cal T}$. Suppose we subdivide ${\cal T}$ into $M = \frac{ 3N^{\cal T} }{ \eps }$ subsegments of equal length, with ${\cal D}$ being their set of endpoints; clearly, $| {\cal D} | = O( \frac{ n^3 }{ \eps^6 } )$. We say that a replenishment policy is ${\cal D}$-restricted when it places orders only at points in ${\cal D}$. The next claim argues that there exists a ${\cal D}$-restricted policy whose time-average cost nearly matches the long-run average cost of the optimal policy ${\cal P}^*$ for our original infinite-horizon instance.

\begin{lemma} \label{lem:D_restrict_for_T}
There exists a ${\cal D}$-restricted policy ${\cal P}$ with $\frac{ F( {\cal P}, {\cal T} ) }{ \| {\cal T} \| } \leq (1 + 31\eps) \cdot F( {\cal P}^* )$. 
\end{lemma}
\begin{proof}
For convenience, we begin by introducing some notation around the  policy ${\cal P}^{\cal T}$. Specifically, for every item $i \in [n]$, let $N_i^{\cal T} = N( {\cal P}^{\cal T}_i, {\cal T})$ be the number of $i$-orders placed along the segment ${\cal T}$, and let $0 = \tau_{i,1} < \tau_{i,2} < \cdots < \tau_{i,N_i^{\cal T}}$ be the sequence of time points where these orders occur. We use $\Delta_{i,1}, \ldots, \Delta_{i,N_i^{\cal T}}$ to denote their durations, namely, $\Delta_{i,1} = \tau_{i,2} - \tau_{i,1}$, $\Delta_{i,2} = \tau_{i,3} - \tau_{i,2}$, so on and so forth. Clearly, $\sum_{\nu \in [N_i^{\cal T}]} \Delta_{i,\nu} = \| {\cal T} \|$. In addition, according to representation~\eqref{eqn:holding_sum_square}, the combined holding cost of this item is given by ${\cal H}_i( {\cal P}^{\cal T}_i, {\cal T}) = H_i \cdot Q( {\cal P}^{\cal T}_i )$, with the convention that $Q( {\cal P}^{\cal T}_i ) = \sum_{\nu \in [N_i^{\cal T}]} \Delta_{i,\nu}^2$.

Now, to define our candidate ${\cal D}$-restricted policy $\hat{\cal P}$, for every item $i \in [n]$, we simply defer each $i$-ordering point $\tau_{i,\nu}$ to $\lceil \tau_{i,\nu} \rceil^{ ({\cal D}) }$, where $\lceil \cdot \rceil^{ ({\cal D}) }$ is an operator that rounds its argument up to the nearest point in ${\cal D}$. It is easy to see that in terms of joint orders, $N( \hat{\cal P}, {\cal T}) \leq N( {\cal P}^{\cal T}, {\cal T})$. Similarly, in terms of $i$-orders, $N( \hat{\cal P}_i, {\cal T}) \leq N( {\cal P}^{\cal T}_i, {\cal T})$ for every item $i \in [n]$. Therefore, the only ingredient where $\hat{\cal P}$ could be more expensive than ${\cal P}^{{\cal T}}$ is its holding cost. 

The important observation is that, letting $\hat{\Delta}_{i,1}, \ldots \hat{\Delta}_{i,N_i^{\cal T}}$ be the sequence of durations between $i$-ordering points with respect to $\hat{\cal P}$, we have $\hat{\Delta}_{i,\nu} \leq {\Delta}_{i,\nu} + \frac{ \| {\cal T} \| }{ M }$, since successive points in ${\cal D}$ are within distance $\frac{ \| {\cal T} \| }{ M }$. We proceed by arguing that $Q( \hat{\cal P}_i ) \leq (1 + \eps) \cdot Q( {\cal P}^{\cal T}_i )$, which would imply in turn that ${\cal H}_i( \hat{\cal P}_i, {\cal T}) \leq (1 + \eps) \cdot {\cal H}_i( {\cal P}^{{\cal T}}_i, {\cal T})$. For this purpose, note that 
\begin{eqnarray}
Q( \hat{\cal P}_i ) & = & \sum_{\nu \in [N_i^{\cal T}]} \hat{\Delta}_{i,\nu}^2 \nonumber \\
& \leq & \sum_{\nu \in [N_i^{\cal T}]} \left( {\Delta}_{i,\nu} + \frac{ \| {\cal T} \| }{ M } \right)^2 \nonumber  \\
& = & \sum_{\nu \in [N_i^{\cal T}]} {\Delta}_{i,\nu}^2 + 2 \cdot \frac{ \| {\cal T} \| }{ M } \cdot \sum_{\nu \in [N_i^{\cal T}]} {\Delta}_{i,\nu} + N_i^{\cal T} \cdot \frac{ \| {\cal T} \|^2 }{ M^2 } \nonumber \\
& = & Q( {\cal P}^{\cal T}_i ) + 2 \cdot \frac{ \| {\cal T} \|^2 }{ M } + N_i^{\cal T} \cdot \frac{ \| {\cal T} \|^2 }{ M^2 } \nonumber \\
& \leq & Q( {\cal P}^{\cal T}_i ) + \left( \frac{ 2\eps }{ 3 N^{\cal T} } + \frac{ \eps^2 }{ 9 N^{\cal T} } \right) \cdot \| {\cal T} \|^2 \label{eqn:lem_D_restrict_for_T_1} \\
& \leq & Q( {\cal P}^{\cal T}_i ) + \frac{ \eps }{ N^{\cal T} } \cdot \| {\cal T} \|^2 \nonumber  \\
& \leq & (1 + \eps) \cdot Q( {\cal P}^{\cal T}_i ) \ . \label{eqn:lem_D_restrict_for_T_2}
\end{eqnarray}
Here, inequality~\eqref{eqn:lem_D_restrict_for_T_1} holds since $M = \frac{ 3N^{\cal T} }{ \eps }$ and  $N_i^{\cal T} \leq N^{\cal T}$. To obtain inequality~\eqref{eqn:lem_D_restrict_for_T_2}, we recall that the  policy ${\cal P}^{{\cal T}}_i$ places $N_i^{\cal T}$ orders of item $i$ along the segment ${\cal T}$. As such, by the convex program bound, $Q( {\cal P}^{\cal T}_i ) \geq \frac{ \| {\cal T} \|^2 }{ N_i^{\cal T} } \geq \frac{ \| {\cal T} \|^2 }{ N^{\cal T} }$; for organizational reasons, a general derivation of this bound is presented in Section~\ref{subsec:useful_bounds} (see Lemma~\ref{lem:convex_bound}). By combining these observations and Lemma~\ref{lem:cheap_subsegment}, it follows that the time-average cost of $\hat{\cal P}$ is
\[ \frac{ F( \hat{\cal P}, {\cal T} ) }{ \| {\cal T} \| } ~~\leq~~ (1 + \eps) \cdot \frac{ F( {\cal P}^{\cal T}, {\cal T} ) }{ \| {\cal T} \| } ~~\leq~~ (1 + 31\eps) \cdot F( {\cal P}^* ) \ . \]
\end{proof}

\paragraph{Summary.} By Lemma~\ref{lem:D_restrict_for_T}, we conclude that rather than directly dealing with our original continuous-time infinite-horizon instance, one can instead compute a near-optimal policy for the discrete-time finite-horizon instance created by subdividing the segment ${\cal T}$ into $O( \frac{ n^3 }{ \eps^6 } )$ uniform-length periods, losing a factor of $1 + O(\eps)$ in optimality. Of course, one issue with this approach is that the length of ${\cal T}$ is generally unknown. However, Lemma~\ref{lem:cheap_subsegment} informs us that $\| {\cal T} \| \in [\frac{ K_0 }{ 16 {\cal M} }, \frac{ 16 {\cal M} }{ \eps^2 H_{\min} }]$. Therefore, losing a factor of $1+\eps$ once again, we can obtain a $(1 + \eps)$-approximation for $\| {\cal T} \|$ by examining $O( \frac{ 1 }{ \eps } \log ( \frac{ {\cal M} }{ \eps K_0 H_{\min} } ))$ options, noting that this quantity is polynomial in $\frac{ 1 }{ \eps }$ as well as in the number of bits required to represent our input, since ${\cal M} = K_0 + n \cdot (K_{\max} + H_{\max})$.
\section{Enhanced Guarantees for the Discrete Formulation} \label{sec:EPTAS}

This section is dedicated to establishing Theorem~\ref{thm:main_EPTAS}, arguing that the discrete-time joint replenishment problem can be approximated within factor $1 + \eps$ of optimal via a deterministic $O( 2^{ \tilde{O}( 1 / \eps^4 ) } \cdot (nT)^{ O(1) } )$-time approach. Toward this objective, Sections~\ref{subsec:useful_bounds} and~\ref{subsec:joint_group_orders} introduce two lower-bounding ideas with respect to single-item policies and elaborate on the applicability of group-fusing, which will be utilized as a dimension reduction tool later on. With these ingredients in place, Section~\ref{subsec:EPTAS_overview}  presents a high-level overview
of our algorithmic approach, followed by a deeper dive into its finer details in Sections~\ref{subsec:easy_scenario}-\ref{subsec:freq_and_avgdense}. Along the way, a number of technical proofs will be delegated to Section~\ref{subsec:add_proofs_discrete}, mainly for ease of exposition.

\subsection{Useful bounds for single-item policies} \label{subsec:useful_bounds}

To streamline certain parts of our analysis, we begin by deriving two useful bounds around discrete-time single-item replenishment policies. For this purpose, suppose that ${\cal P}$ is such a policy, defined across $T$ time periods. Letting $N( {\cal P} )$ be the number of orders placed by ${\cal P}$, with $\Delta_1, \ldots, \Delta_{N({\cal P})}$ being their  durations, we remind the reader that the operational cost of this policy can be written as $C( {\cal P} ) = K \cdot N({\cal P}) + H \cdot Q({\cal P})$, with the convention that $Q( {\cal P} ) = \sum_{\nu \in [N( {\cal P} )]} \Delta_{\nu}^2$.

\paragraph{The EOQ-based bound.} The first bound we wish to highlight is obtained by comparing the cost of ${\cal P}$ to that of an optimal policy for the economic order quantity (EOQ) model of the item in question.

\begin{lemma} \label{lem:EOQ_bound}
$C( {\cal P} ) \geq 2T  \sqrt{KH}$.
\end{lemma}
\begin{proof}
To determine a lower bound on $C( {\cal P} )$, let us view ${\cal P}$ as if this policy operates in continuous time across the interval $[0,T]$. From this perspective, by duplicating the latter interval, we create a feasible dynamic policy for the  continuous-time infinite-horizon EOQ model, whose long-run average cost is $\frac{ C( {\cal P} ) }{ T }$. However, following  Section~\ref{subsec:model_definition}, we know that this model admits a periodic optimal policy, attaining a long-run average cost of $2\sqrt{KH}$, as stated in Claim~\ref{clm:EOQ_properties}. Therefore, $\frac{ C( {\cal P} ) }{ T } \geq 2\sqrt{KH}$, as desired.
\end{proof}

\paragraph{The convex program bound.} Our second lower-bounding idea is not directly concerned with the overall cost $C( {\cal P} )$, but rather with the sum of squared durations, $Q( {\cal P} ) = \sum_{\nu \in [N( {\cal P} )]} \Delta_{\nu}^2$. Here, we employ basic convexity arguments to bound this quantity in terms of the number of underlying periods $T$ and the number of orders $N( {\cal P} )$ placed by ${\cal P}$.

\begin{lemma} \label{lem:convex_bound}
$Q( {\cal P} ) \geq \frac{ T^2 }{ N( {\cal P} ) }$.
\end{lemma}
\begin{proof}
Letting $N = N( {\cal P} )$, the important observation is that the vector $(\Delta_1, \ldots, \Delta_N)$ of order durations is a feasible solution to the following convex optimization problem:
\begin{equation} \label{eqn:convex_prog_Q} \tag{CP}
\begin{array}{ll}
\min & {\displaystyle \sum_{\nu \in [N]} x_{\nu}^2} \\
\text{s.t.} & {\displaystyle \sum_{\nu \in [N]} x_{\nu} = T} \\
& x \in \bbR^{ N }_+ 
\end{array}    
\end{equation}
In what follows, we show that~\eqref{eqn:convex_prog_Q} has a unique optimal solution, given by $x^*_1 = \cdots = x^*_{N} = \frac{ T }{ N }$, implying that $\opt\eqref{eqn:convex_prog_Q} = \frac{ T^2 }{ N }$.  Therefore, $Q( {\cal P} ) = \sum_{\nu \in [N]} {\Delta}_{\nu}^2 \geq \frac{ T^2 }{ N }$. 

To argue about the optimality and uniqueness of $x^*$, suppose that $\hat{x} \in \bbR^{ N }_+$ is an optimal solution, with $\hat{x} \neq x^*$. Then, since $\sum_{\nu \in [N]} \hat{x}_{\nu} = T$, there exists a pair of coordinates $\nu_1 \neq \nu_2$ for which $\hat{x}_{\nu_1} > \frac{ T }{ N }$ and $\hat{x}_{\nu_2} < \frac{ T }{ N }$. Letting $\delta = \min \{ \hat{x}_{\nu_1} - \frac{ T }{ N }, \frac{ T }{ N } - \hat{x}_{\nu_2} \} > 0$, consider the vector $\tilde{x} \in \bbR^{ N }_+$, given by
\[ \tilde{x}_{\nu} ~~=~~ \begin{dcases}
\hat{x}_{\nu}, \qquad  & \nu \in [N] \setminus \{ \nu_1, \nu_2 \} \\
\hat{x}_{\nu_1} - \delta, \qquad  & \nu = \nu_1 \\
\hat{x}_{\nu_2} + \delta, \qquad  & \nu = \nu_2 
\end{dcases} \]
Clearly, $\tilde{x}$ is a feasible solution to~\eqref{eqn:convex_prog_Q}, and  we arrive at a contradiction to the optimality of $\hat{x}$, since 
\begin{eqnarray*}
\sum_{\nu \in [N]} \hat{x}_{\nu}^2 - \sum_{\nu \in [N]} \tilde{x}_{\nu}^2 & = & \hat{x}_{\nu_1}^2 - ( \hat{x}_{\nu_1} - \delta )^2 + \hat{x}_{\nu_2}^2 - ( \hat{x}_{\nu_2} + \delta )^2 \\
& = & 2 \delta \cdot ( \hat{x}_{\nu_1} - \hat{x}_{\nu_2} - \delta ) \\
& \geq & 2\delta^2 \\
& > & 0 \ ,
\end{eqnarray*}
where the first and second inequalities respectively hold since $\delta = \min \{ \hat{x}_{\nu_1} - \frac{ T }{ N }, \frac{ T }{ N } - \hat{x}_{\nu_2} \} \leq \frac{ \hat{x}_{\nu_1} - \hat{x}_{\nu_2} }{ 2 }$ and $\delta > 0$.
\end{proof}

\subsection{Ratio-groups and group-fusing policies } \label{subsec:joint_group_orders}

\paragraph{Item classification.} For every item $i \in [n]$, we begin by rounding its ordering cost $K_i$ up to the nearest integer power of $1 + \eps$; the same goes for its holding cost, $H_i$. It is easy to verify that, as a result of these alterations, the operational cost of any policy blows-up by a factor of at most $1 + \eps$. To avoid excessive notation, $\{ K_i \}_{i \in [n]}$ and $\{ H_i \}_{i \in [n]}$ will be kept as our ordering and holding costs. Next, we partition the underlying set of items into ratio-groups, ${\cal G}_1, \ldots, {\cal G}_L$, such that within each group ${\cal G}_{\ell}$, all items are associated with the same ordering-to-holding cost ratio $\frac{ K_i }{ H_i }$. The common ratio of ${\cal G}_{\ell}$-items will be designated by $\rho_{\ell}$, noting that this parameter must be an integer power of $1 + \eps$, due to our cost rounding operations.

\paragraph{Jointly ordering groups.} We say that a replenishment policy ${\cal P}$ is group-fusing when, for every $\ell \in [L]$, all items in ${\cal G}_{\ell}$ are jointly ordered throughout the entire planning horizon. In other words, at every time period $t \in [T]$, either all ${\cal G}_{\ell}$-items are ordered or none are ordered. The next structural property, originally observed by \citet[Sec.~2.2]{Segev14}, argues that by focusing our attention on group-fusing policies, we are not sacrificing optimality. For completeness, we provide a clean proof of this claim.

\begin{lemma} \label{lem:opt_group_fusing}
There exists a group-fusing optimal replenishment policy.
\end{lemma}
\begin{proof}
With respect to a replenishment policy ${\cal P}$, the ratio-group ${\cal G}_{\ell}$ is called split when there is at least one time period where a non-empty proper subset of the items in ${\cal G}_{\ell}$ is ordered. The number of ratio-groups split by ${\cal P}$ will be referred to as $s( {\cal P} )$, noting that $s( {\cal P} ) = 0$ is equivalent to ${\cal P}$ being group-fusing.

In terms of this definition, over all optimal policies, let ${\cal P}^*$ be one whose number of split ratio-groups $s( \cdot )$ is minimal; we claim that $s( {\cal P}^* ) = 0$. Otherwise, let ${\cal G}_{\ell}$ be some group split by ${\cal P}^*$, and let $i^* \in {\cal G}_{\ell}$ be an item for which $\rho_{\ell} \cdot N( {\cal P}^*_i ) + Q( {\cal P}^*_i )$ is minimized across all ${\cal G}_{\ell}$-items. Suppose we define a new policy $\hat{\cal P}$, which is identical to ${\cal P}^*$, except for utilizing the policy ${\cal P}^*_{i^*}$ for all items in ${\cal G}_{\ell}$. Clearly, $J( \hat{\cal P} ) = J( {\cal P}^* )$, since we are not creating new joint orders. Similarly, $C_i( \hat{\cal P}_i ) = C_i( {\cal P}^*_i )$ for every item $i \notin {\cal G}_{\ell}$, since its replenishment policy remains unchanged. However, for every item $i \in {\cal G}_{\ell}$, 
\begin{eqnarray*}
C_i( \hat{\cal P}_i ) & = & C_i( {\cal P}^*_{i^*} )  \\
& = & K_i \cdot N( {\cal P}^*_{i^*} ) + H_i \cdot Q( {\cal P}^*_{i^*} )  \\
& = & H_i \cdot ( \rho_{\ell} \cdot N( {\cal P}^*_{i^*} ) + Q( {\cal P}^*_{i^*} ) )  \\
& \leq & H_i \cdot ( \rho_{\ell} \cdot N( {\cal P}^*_i ) + Q( {\cal P}^*_i ) )  \\
& = & C_i( {\cal P}^*_i ) \ ,  
\end{eqnarray*}
where the sole  inequality above  follows from how item $i^*$ is defined. Consequently, $\hat{\cal P}$ must be an optimal policy as well, and since ${\cal G}_{\ell}$ is now jointly ordered, $s( \hat{\cal P} ) = s( {\cal P}^* ) - 1$, contradicting the choice of ${\cal P}^*$ as an optimal policy of minimum $s( \cdot )$ value.
\end{proof}

\paragraph{Simplified notation.} In the remainder of this section, we will be focusing on group-fusing replenishment policies. Therefore, rather than working with $1, \ldots, n$ as our collection of items, we will unify those residing within each ratio-group ${\cal G}_{\ell}$ into a representative super-item, designated by $\ell$. For ease of notation, the ordering and holding costs of this item will be $K_{\ell} = \sum_{i \in {\cal G}_{\ell}} K_i$ and $H_{\ell} = \sum_{i \in {\cal G}_{\ell}} H_i$. Clearly, $\frac{ K_{\ell} }{ H_{\ell} } = \rho_{\ell}$, meaning in particular that this ratio is an integer power of $1 + \eps$.

\subsection{Technical overview} \label{subsec:EPTAS_overview}

In what follows, we provide a high-level outline of our approach for computing a near-optimal approximate policy with respect to the simplified instance constructed in Section~\ref{subsec:joint_group_orders}. To this end, letting ${\cal P}^*$ be an optimal policy in this context, we begin by guessing the number of joint orders $N^* = N( {\cal P}^* )$ placed by ${\cal P}^*$ across the entire planning horizon, noting that there are only $T$ candidate values to be tested. Next, we say that our instance falls into the ``easy'' scenario when relatively few joint orders are placed, specifically meaning that  $N^* < \frac{ 1 }{ \eps^4 }$. The opposite case, where $N^* \geq \frac{ 1 }{ \eps^4 }$, will be referred to as the ``difficult'' scenario. 

In Section~\ref{subsec:easy_scenario},  we employ efficient enumeration ideas to compute an $\eps$-optimal replenishment policy for easy instances, in $O( 2^{ \tilde{O}( 1 / \eps^4 ) } \cdot (nT)^{ O(1) } )$  time. As an aside, it is important to point out that a naive approach, where one enumerates over all $O( T^{ N^* } )$ configurations for placing joint orders, does not lead to an EPTAS, which is what we are looking for in Theorem~\ref{thm:main_EPTAS}. Subsequently, Sections~\ref{subsec:rare_grid} and~\ref{subsec:freq_and_avgdense} will be dedicated to dealing with  difficult instances, on which we proceed to elaborate next.

\paragraph{The difficult scenario: Items types.} As our starting point for addressing this scenario, let us partition the set of items $[L]$ into four types, depending on their ordering-to-holding cost ratio $\rho_{\ell} = \frac{ K_{\ell} }{ H_{\ell} }$ and on their yet-unknown number of orders $N_{\ell}^* = N( {\cal P}^*_{\ell} )$. Specifically, we say that item $\ell \in [L]$ is rare when $\sqrt{ \rho_{\ell} } \geq \frac{ T }{ \eps^2 N^* }$. On the other hand, item $\ell$ is said to be frequent when $\sqrt{ \rho_{\ell} } \leq \frac{ 2\eps T }{ N^* }$. All other items will be referred to as being average, with $\rho_{\ell} \in (\frac{ 2\eps T }{ N^* }, \frac{ T }{ \eps^2 N^* })$. As explained in Section~\ref{subsec:joint_group_orders}, the cost ratios $\{ \rho_{\ell} \}_{\ell \in [L]}$ are all integer powers of $1+\eps$, meaning that there are only $O( \frac{ 1 }{ \eps } \log \frac{ 1 }{ \eps } )$ average items. For any such item $\ell$, we guess whether $N_{\ell}^* \geq \eps^2 N^*$ or not, for which the required number of guesses is only $2^{ \tilde{O}( 1/\eps ) }$. Items of the former type will be called average-dense, whereas those of the latter type are said to be average-sparse. We denote the collections of rare, frequent, average-dense, and average-sparse items by ${\cal R}$, ${\cal F}$,  ${\cal A}_{\mydense}$, and ${\cal A}_{\mysparse}$, respectively. 

\paragraph{The difficult scenario: Type-specific policies.} Quite surprisingly, our approach consists of independently proposing one replenishment policy for rare and average-sparse items, ${\cal P}^{ {\cal R} \uplus {\cal A}_{\mysparse}}$, and another policy for frequent and average-dense items, ${\cal P}^{ {\cal F} \uplus {\cal A}_{\mydense}}$. Specifically, in Section~\ref{subsec:rare_grid}, we devise a deterministic $O( (nT)^{ O(1) } )$-time algorithm for computing a policy ${\cal P}^{ {\cal R} \uplus {\cal A}_{\mysparse}}$ whose operational cost is
\[ F( {\cal P}^{ {\cal R} \uplus {\cal A}_{\mysparse}} ) ~~\leq~~ 2\eps K_0 N^* + (1 + 9 \eps) \cdot \sum_{\ell \in {\cal R} \uplus {\cal A}_{\mysparse} } C_{\ell}( {\cal P}_{\ell}^* ) \ . \] 
Then, in Section~\ref{subsec:freq_and_avgdense}, we propose a deterministic $O( 2^{ O(1/\eps^4) } \cdot (nT)^{ O(1) } )$-time algorithm for constructing a policy ${\cal P}^{ {\cal F} \uplus {\cal A}_{\mydense}}$, such that
\[ F( {\cal P}^{ {\cal F} \uplus {\cal A}_{\mydense}} ) ~~\leq~~ (1 + \eps) \cdot K_0 N^* + (1 + 17 \eps) \cdot \sum_{\ell \in {\cal F} \uplus {\cal A}_{\mydense} } C_{\ell}( {\cal P}_{\ell}^* ) \ . \] 
Letting ${\cal P}^{ [L] }$ be the policy obtained by gluing ${\cal P}^{ {\cal R} \uplus {\cal A}_{\mysparse}}$ and ${\cal P}^{ {\cal F} \uplus {\cal A}_{\mydense}}$ together, it follows that our resulting operational cost is
\[ F( {\cal P}^{ [L] } ) ~~\leq~~ (1 + 3\eps) \cdot K_0 N^* + (1 + 17 \eps) \cdot \sum_{\ell \in [L] } C_{\ell}( {\cal P}_{\ell}^* ) ~~\leq~~ (1 + 17 \eps) \cdot F( {\cal P}^* ) \ . \]

\subsection{The easy scenario \texorpdfstring{$\bs{(N^* < \frac{ 1 }{ \eps^4 })}$}{}} \label{subsec:easy_scenario}

\paragraph{The enumeration procedure.} With respect to the unknown optimal policy ${\cal P}^*$, let $1 = \tau_1^* < \tau_2^* < \cdots < \tau_{N^*}^*$ be the sequence of time periods where joint order occur. In addition, let $\Delta_1^*, \ldots, \Delta_{N^*}^*$ be the durations of these orders, namely, $\Delta_1^* = \tau_2^* - \tau_1^*$, $\Delta_2^* = \tau_3^* - \tau_2^*$, so on and so forth. With this notation, our algorithm proceeds as follows:
\begin{enumerate}
    \item \label{item:enumeration_1} 
    {\em Guessing}: For every $\nu \in [N^*]$, we guess an  over-estimate $\tilde{\Delta}_{\nu}$ for the order duration $\Delta^*_{\nu}$, such that 
    \begin{equation} \label{eqn:relation_estimates}
    \Delta^*_{\nu} ~~\leq~~ \tilde{\Delta}_{\nu} ~~\leq~~ (1 + \eps) \cdot \Delta^*_{\nu} \ .
    \end{equation}
    Since each duration $\Delta^*_{\nu}$ is bounded within $[1,T]$, and since the easy scenario corresponds to $N^* < \frac{ 1 }{ \eps^4 }$, the combined number of required guesses for $\tilde{\Delta} = (\tilde{\Delta}_{\nu})_{ \nu \in [N^*] }$ is
    \[ O \left( \left(\frac{ 1 }{ \eps } \log T \right)^{ N^* } \right) ~~=~~ O \left( \left(\frac{ 1 }{ \eps } \log T \right)^{ 1/\eps^4 } \right) ~~=~~ O \left( T \cdot 2^{ \tilde{O}( 1/\eps^4 )} \right) \ . \]    

    \item \label{item:enumeration_2} {\em Placing joint orders}: Given $\tilde{\Delta}_1, \ldots, \tilde{\Delta}_{N^*}$, we reverse-engineer the placement of joint orders, which will be occurring at the sequence of time periods $1 = \tilde{\tau}_1 < \tilde{\tau}_2 < \cdots < \tilde{\tau}_{N^*}$, where $\tilde{\tau}_2 = \tilde{\tau}_1 + \tilde{\Delta}_1$, $\tilde{\tau}_3 = \tilde{\tau}_2 + \tilde{\Delta}_2$, so on and so forth. It is worth mentioning that, in light of relation~\eqref{eqn:relation_estimates}, we have $\sum_{\nu \in [N^*]} \tilde{\Delta}_{\nu} \geq \sum_{\nu \in [N^*]} \Delta^*_{\nu} = T$, meaning that our policy may extend beyond $T$ periods. Of course, truncating to exactly $T$ periods will only be cost-saving.

    \item \label{item:enumeration_3} {\em Placing individual orders}: Restricted to these joint orders, for each item $\ell \in [L]$, we separately compute an optimal policy $\tilde{\cal P}_{\ell}$, i.e., one minimizing $C_{\ell}( \cdot ) = K_{\ell} \cdot N(\cdot) + H_{\ell} \cdot Q(\cdot)$. This setting can be viewed as a structured special case of the discrete-time single-item lot sizing problem, whose  optimal policy can easily be identified in polynomial time (for example, via the Wagner-Whitin algorithm \citeyearpar{WagnerW58}).
\end{enumerate}

\paragraph{Analysis.} We conclude this section by arguing that $\tilde{\cal P} = ( \tilde{\cal P}_1, \ldots, \tilde{\cal P}_L)$ is a near-optimal policy. Noting that this policy matches ${\cal P}^*$ in terms of joint orders, the quintessential question would be: Given our order-spacing method (see  items~\ref{item:enumeration_1} and~\ref{item:enumeration_2} above), can we still compare the best-possible lot sizing cost of each item to its analogous cost with respect to ${\cal P}^*$?

\begin{lemma} \label{lem:easy_final_approx}
$F( \tilde{\cal P} ) \leq (1 + 3\eps) \cdot F( {\cal P}^* )$. 
\end{lemma}
\begin{proof}
We first observe that $\tilde{\cal P}$ places precisely the same number $N^* = N( {\cal P}^* )$ of joint orders as the optimal policy ${\cal P}^*$, meaning that $J( \tilde{\cal P} ) = J( {\cal P}^* )$. The crux of our proof resides in showing that in terms of marginal costs, $C_{\ell}( \tilde{\cal P}_{\ell} ) \leq (1 + 3\eps) \cdot C_{\ell}( {\cal P}^*_{\ell} )$ for every item $\ell \in [L]$. To establish this claim, it suffices to present a feasible policy $\hat{\cal P}_{\ell}$ for the lot sizing problem of each item $\ell$ whose cost is at most $(1 + 3\eps) \cdot C_{\ell}( {\cal P}^*_{\ell} )$. Specifically, the latter policy will satisfy $N( \hat{\cal P}_{\ell} ) = N( {\cal P}_{\ell}^* )$ and $Q( \hat{\cal P}_{\ell} ) \leq (1 + 3\eps) \cdot Q( {\cal P}_{\ell}^* )$. 

For this purpose, suppose that the sequence of $\ell$-ordering points with respect to ${\cal P}^*_{\ell}$ is given by $1 = \tau_{\ell,1}^* < \tau_{\ell,2}^* < \cdots < \tau_{\ell,N_{\ell}^*}^*$, with the convention that $N_{\ell}^* = N( {\cal P}^*_{\ell} )$; this sequence is obviously a subsequence of $1 = \tau_1^* < \tau_2^* < \cdots < \tau_{N^*}^*$. Our candidate policy $\hat{\cal P}_{\ell}$ places $\ell$-orders at the time periods $1 = \tilde{\tau}_{\ell,1} < \tilde{\tau}_{\ell,2} < \cdots < \tilde{\tau}_{\ell,N_{\ell}^*}$. As a result, $N( \hat{\cal P}_{\ell} ) = N_{\ell}^* = N( {\cal P}^*_{\ell} )$, and it remains to show that $Q( \hat{\cal P}_{\ell} ) \leq (1 + 3\eps) \cdot Q( {\cal P}^*_{\ell} )$. This relation follows by observing that
\begin{eqnarray*}
Q( \hat{\cal P}_{\ell} ) & = & \sum_{\nu \in [N_{\ell}^*]} ( \tilde{\tau}_{\ell,\nu+1} - \tilde{\tau}_{\ell,\nu} )^2  \\
& \leq & (1 + \eps)^2 \cdot  \sum_{\nu \in [N_{\ell}^*]} \left( {\tau}_{\ell,\nu+1}^* - {\tau}_{\ell,\nu}^* \right)^2  \\
& < & (1 + 3\eps ) \cdot Q( {\cal P}_{\ell}^* ) \ ,
\end{eqnarray*}
where the first inequality holds since, according to~\eqref{eqn:relation_estimates}, the $\tilde{\Delta}$-duration of each joint order between $\tilde{\tau}_{\ell,\nu}$ and $\tilde{\tau}_{\ell,\nu+1}$ is longer by a factor of at most $1 + \eps$ compared to the $\Delta^*$-duration of its corresponding order between ${\tau}^*_{\ell,\nu}$ and ${\tau}^*_{\ell,\nu+1}$.
\end{proof}

% \Danny{Previous explanations, maybe relevant to a subsequent proof: \\
% Here, inequality~????? holds since there are  most $N^* - 2$ joint orders between $\tilde{\tau}_{\ell,\nu}$ and $\tilde{\tau}_{\ell,\nu+1}$, and since the durations of these orders in terms of $\tilde{\Delta}$ and $\Delta^*$ differ by at most $\frac{ \eps^5 T }{N^*}$, according to~\eqref{eqn:relation_estimates}. Inequalities~????? and~???? are respectively obtained by recalling that $\sum_{\nu \in [N_{\ell}^*]} ( {\tau}_{\ell,\nu+1}^* - {\tau}_{\ell,\nu}^* ) = T$ and that $N_{\ell}^* \leq N^* < \frac{ 1 }{ \eps^4}$, due to currently considering the easy scenario. Finally, to derive inequality~?????, we observe that ${\cal P}_{\ell}^*$ places $N_{\ell}^*$ orders across a  planning horizon of length $T$. Therefore, by the convex program bound in Lemma~\ref{lem:convex_bound}, we have $Q( {\cal P}_{\ell}^* ) \geq \frac{ T^2 }{ N_{\ell}^* } \geq \frac{ T^2 }{ N^* } > \eps^4 T^2$.}

\subsection{The difficult scenario \texorpdfstring{$\bs{(N^* \geq \frac{ 1 }{ \eps^4 })}$}{}: Rare and average-sparse items} \label{subsec:rare_grid}

Moving on to consider the difficult scenario, as outlined in Section~\ref{subsec:EPTAS_overview}, our first objective is to deterministically compute a replenishment policy ${\cal P}^{ {\cal R} \uplus {\cal A}_{\mysparse}}$ for rare and average-sparse items, with an operational cost of 
\[ F( {\cal P}^{ {\cal R} \uplus {\cal A}_{\mysparse}} ) ~~\leq~~ 2\eps K_0 N^* + (1 + 9 \eps) \cdot \sum_{\ell \in {\cal R} \uplus {\cal A}_{\mysparse} } C_{\ell}( {\cal P}_{\ell}^* ) \ . \] 

\paragraph{The evenly-spaced grid $\bs{\cal T}$.} To this end, let us temporarily leave out average-sparse items, and focus our attention on rare ones. Suppose we create a grid ${\cal T} \subseteq [T]$ of  time points, evenly-spaced by $\Gamma = \left\lceil \frac{ T }{ \eps N^* } \right\rceil$ periods, where ${\cal T} = \left\{ 1, 1 + \Gamma, 1 + 2 \Gamma, \ldots \right\}$. We say that a replenishment policy for the collection of rare items is ${\cal T}$-restricted when it places orders only at ${\cal T}$-points. Clearly, since $|{\cal T}| \leq 2\eps N^*$,  the joint ordering cost of any such policy is at most a $2\eps$-fraction of the analogous cost in the optimal policy ${\cal P}^*$.

\paragraph{$\bs{\cal T}$-restricted policies for rare items?} In what follows, we prove that in spite of their highly structured nature, ${\cal T}$-restricted policies can compete against ${\cal P}^*$ when it comes to the marginal cost of every rare item.

\begin{lemma} \label{lem:grid_rare}
There exists a ${\cal T}$-restricted policy ${\cal P}^{\cal R}$, where $C_{\ell}( {\cal P}_{\ell}^{\cal R} ) \leq (1 + 9\eps) \cdot C_{\ell}( {\cal P}_{\ell}^* )$ for every $\ell \in {\cal R}$.
\end{lemma}
\begin{proof}
To define the desired policy ${\cal P}^{\cal R}$, we first place joints orders at all points of the grid ${\cal T}$. Next, to determine the replenishment policy ${\cal P}^{\cal R}_{\ell}$ of each item $\ell \in {\cal R}$, we consider two cases, depending on the relation between $N_{\ell}^* = N( {\cal P}_{\ell}^* )$ and $\frac{ 1 }{ \eps }$.

\paragraph{Case 1: $\bs{N_{\ell}^* \leq \frac{ 1 }{ \eps }}$.} Suppose that the sequence of $\ell$-ordering points with respect to ${\cal P}^*_{\ell}$ is given by $1 = \tau_{\ell,1}^* < \tau_{\ell,2}^* < \cdots < \tau_{\ell,N_{\ell}^*}^*$. Then, our policy ${\cal P}_{\ell}^{\cal R}$ will place $\ell$-orders at the time points $1 = \lceil \tau_{\ell,1}^* \rceil^{ ({\cal T}) } \leq \lceil \tau_{\ell,2}^* \rceil^{ ({\cal T}) } \leq \cdots \leq \lceil \tau_{\ell,N_{\ell}^*}^* \rceil^{ ({\cal T}) }$, where $\lceil \cdot \rceil^{ ({\cal T}) }$ is an operator that rounds its argument up to the nearest point in ${\cal T}$. We proceed by showing that $C_{\ell}( {\cal P}_{\ell}^{\cal R} ) \leq (1 + 8\eps) \cdot C_{\ell}( {\cal P}_{\ell}^* )$. For this purpose, according to the definition of ${\cal P}_{\ell}^{\cal R}$, its number of $\ell$-orders is clearly upper-bounded by that of ${\cal P}_{\ell}^*$. As far as holding costs are concerned, we observe that ${\cal H}_{\ell}( {\cal P}_{\ell}^{\cal R} ) \leq (1 + 8\eps) \cdot {\cal H}_{\ell}( {\cal P}_{\ell}^* )$, since
\begin{eqnarray}
Q( {\cal P}^{\cal R}_{\ell} ) & = & \sum_{\nu \in [N_{\ell}^*]} ( \lceil \tau_{\ell,\nu+1}^* \rceil^{ ({\cal T}) } - \lceil \tau_{\ell,\nu}^* \rceil^{ ({\cal T}) } )^2 \nonumber \\
& \leq & \sum_{\nu \in [N_{\ell}^*]} \left( {\tau}_{\ell,\nu+1}^* + \Gamma - {\tau}_{\ell,\nu}^* \right)^2 \label{eqn:lem_grid_rare_1} \\
& = & Q( {\cal P}_{\ell}^* ) + 2\Gamma T + N_{\ell}^* \Gamma^2 \label{eqn:lem_grid_rare_2} \\
& \leq & Q( {\cal P}_{\ell}^* ) + \frac{ 8T^2 }{ \eps^2 N^* } \label{eqn:lem_grid_rare_3} \\
& \leq & Q( {\cal P}_{\ell}^* ) + 8 \eps^2 T^2 \label{eqn:lem_grid_rare_4}  \\
& \leq & (1 + 8\eps ) \cdot Q( {\cal P}_{\ell}^* ) \ . \label{eqn:lem_grid_rare_5}
\end{eqnarray}
Here, inequality~\eqref{eqn:lem_grid_rare_1} holds since $\lceil \tau_{\ell,\nu}^* \rceil^{ ({\cal T}) } \geq \tau_{\ell,\nu}^*$ and since $\lceil \tau_{\ell,\nu+1}^* \rceil^{ ({\cal T}) } \leq \tau_{\ell,\nu+1}^* + \Gamma$. Inequalities~\eqref{eqn:lem_grid_rare_2} and~\eqref{eqn:lem_grid_rare_3} are obtained by recalling that $\sum_{\nu \in [N_{\ell}^*]} ( {\tau}_{\ell,\nu+1}^* - {\tau}_{\ell,\nu}^* ) = T$, $\Gamma = \lceil \frac{ T }{ \eps N^* } \rceil \leq \frac{ 2T }{ \eps N^* }$, and $N_{\ell}^* \leq N^*$. Inequality~\eqref{eqn:lem_grid_rare_4} follows by noting that $N^* \geq \frac{ 1 }{ \eps^4 }$, as we are currently considering the difficult scenario. Lastly, to better understand where inequality~\eqref{eqn:lem_grid_rare_5} is coming from, recall that ${\cal P}_{\ell}^*$ places $N_{\ell}^*$ orders across a  planning horizon of length $T$. Therefore, by  Lemma~\ref{lem:convex_bound}, we have $Q( {\cal P}_{\ell}^* ) \geq \frac{ T^2 }{ N_{\ell}^* } \geq \eps T^2$, where the last transition follows from the hypothesis of case~1, stating that $N_{\ell}^* \leq \frac{ 1 }{ \eps }$.

\paragraph{Case 2: $\bs{N_{\ell}^* > \frac{ 1 }{ \eps }}$.} Let $\psi_{\ell}$ be the unique integer for which $(\psi_{\ell}-1) \cdot \Gamma \leq \sqrt{ \rho_{\ell} } < \psi_{\ell}  \Gamma$. By this definition, 
\begin{equation} \label{eqn:UB_psi_ell}
\psi_{\ell} ~~>~~ \frac{ \sqrt{ \rho_{\ell} } }{ \Gamma } ~~\geq~~ \frac{ T / (\eps^2 N^* ) }{ 2 T / (\eps N^*) } ~~=~~ \frac{ 1 }{ 2\eps } \ .    
\end{equation}
The second inequality is obtained by noting that item $\ell$ is rare, and therefore, $\sqrt{ \rho_{\ell} } \geq \frac{ T }{ \eps^2 N^* }$. In addition, $\Gamma \leq \frac{ 2T }{ \eps N^* }$, as observed when proving case~1, independently of its hypothesis. Given $\psi_{\ell}$, our policy ${\cal P}_{\ell}^{\cal R}$ will place $\ell$-orders at the time points $1, 1 + \psi_{\ell}  \Gamma, 1 + 2\psi_{\ell}  \Gamma, \ldots$, so on and so forth. Consequently, we arrive at an operational cost of
\begin{eqnarray*}
C_{\ell}( {\cal P}_{\ell}^{\cal R} ) & = & K_{\ell} \cdot N( {\cal P}_{\ell}^{\cal R} ) + H_{\ell} \cdot Q( {\cal P}_{\ell}^{\cal R} )  \\
& \leq & K_{\ell} \cdot \left\lceil \frac{ T }{ \psi_{\ell}\Gamma } \right\rceil + H_{\ell} \cdot \left\lceil \frac{ T }{ \psi_{\ell}\Gamma } \right\rceil \cdot \left( \psi_{\ell}\Gamma\right)^2  \\
& \leq & \underbrace{ \frac{ K_{\ell}T }{ \psi_{\ell}\Gamma }  + H_{\ell} T \psi_{\ell}\Gamma }_{ \mathrm{(I)} } + \underbrace{ K_{\ell} + H_{\ell} \cdot \left( \psi_{\ell} \Gamma \right)^2 }_{ \mathrm{(II)} }  \\
& \leq & (1 + 9\eps) \cdot C_{\ell}( {\cal P}_{\ell}^* ) \ , 
\end{eqnarray*}
where the last inequality follows from the next auxiliary claim, whose proof is provided in Section~\ref{subsec:add_proofs_discrete}. 

\begin{claim} \label{clm:aux_grid_rare}
$\mathrm{(I)} \leq (1 + 4\eps) \cdot C_{\ell}( {\cal P}_{\ell}^* )$ and $\mathrm{(II)} \leq 5\eps \cdot C_{\ell}( {\cal P}_{\ell}^* )$.
\end{claim} 
\end{proof}

\paragraph{$\bs{\cal T}$-restricted policies for average-sparse items?} Along the same lines, we argue that ${\cal T}$-restricted policies can compete against the optimal policy ${\cal P}^*$ in terms of minimizing the marginal cost of every average-sparse item. Actually, for the purpose of deriving Lemma~\ref{lem:grid_for_sparse} below, we will only be exploiting sparsity. To avoid repetitive arguments, since the proof of this result is rather similar to that of case~1 in Lemma~\ref{lem:grid_rare}, we provide its finer details in Section~\ref{subsec:add_proofs_discrete}. 

\begin{lemma} \label{lem:grid_for_sparse}
There exists a ${\cal T}$-restricted policy ${\cal P}^{{\cal A}_{\mysparse}}$, where $C_{\ell}( {\cal P}_{\ell}^{\cal R} ) \leq (1 + 8\eps) \cdot C_{\ell}( {\cal P}_{\ell}^* )$ for every $\ell \in {\cal A}_{\mysparse}$.
\end{lemma}

\paragraph{The resulting policy.} With these ingredients in place, our policy for rare and average-sparse items is pretty straightforward: We construct the grid ${\cal T}$ and separately compute optimal ${\cal T}$-restricted policies, ${\cal P}^{\cal R}$ and ${\cal P}^{{\cal A}_{\mysparse}}$, in the sense of minimizing $\sum_{\ell \in {\cal R}} C_{\ell}( {\cal P}_{\ell}^{\cal R} )$ and $\sum_{\ell \in {\cal A}_{\mysparse}} C_{\ell}( {\cal P}^{{\cal A}_{\mysparse}}_{\ell} )$, respectively. To this end, one can view ${\cal T}$ as the set of time periods where orders are allowed to be placed, independently solving a discrete-time single-item lot sizing problem for each item $\ell \in {\cal R} \uplus {\cal A}_{\mysparse}$, as explained in item~\ref{item:enumeration_3} of Section~\ref{subsec:easy_scenario}. By gluing these two policies together, we end up with a single policy ${\cal P}^{ {\cal R} \uplus {\cal A}_{\mysparse}}$ whose combined cost is  
\begin{eqnarray*}
F( {\cal P}^{ {\cal R} \uplus {\cal A}_{\mysparse}} ) & = & K_0 \cdot |{\cal T}| + \sum_{\ell \in {\cal R}} C_{\ell}( {\cal P}_{\ell}^{\cal R} ) + \sum_{\ell \in {\cal A}_{\mysparse}} C_{\ell}( {\cal P}^{{\cal A}_{\mysparse}}_{\ell} ) \\
& \leq &2\eps K_0 N^* + (1 + 9 \eps) \cdot \sum_{\ell \in {\cal R} \uplus {\cal A}_{\mysparse} } C_{\ell}( {\cal P}_{\ell}^* ) \ ,    
\end{eqnarray*}
where the inequality above follows by recalling that $|{\cal T}| \leq 2\eps N^*$, and by plugging in Lemmas~\ref{lem:grid_rare} and~\ref{lem:grid_for_sparse}.

\subsection{The difficult scenario \texorpdfstring{$\bs{(N^* \geq \frac{ 1 }{ \eps^4 })}$}{}: Frequent and average-dense items} \label{subsec:freq_and_avgdense}

Following the global plan described in Section~\ref{subsec:EPTAS_overview}, we turn our attention to devising a dynamic programming approach for treating frequent and average-dense items. These ideas will enable us to deterministically compute a replenishment policy ${\cal P}^{ {\cal F} \uplus {\cal A}_{\mydense}}$ for such items, with an operational cost of  
\[ F( {\cal P}^{ {\cal F} \uplus {\cal A}_{\mydense}} ) ~~\leq~~ (1 + \eps) \cdot K_0 N^* + (1 + 17 \eps) \cdot \sum_{\ell \in {\cal F} \uplus {\cal A}_{\mydense} } C_{\ell}( {\cal P}_{\ell}^* ) \ . \] 
For convenience, we assume without loss of generality that $\frac{ 1 }{ \eps }$ takes an integer value. 

\paragraph{Including frequent items in all orders?} By definition of average-dense items, we know that $N_{\ell}^* \geq \eps^2 N^*$ for any such item, meaning that it is ordered at an $\Omega( \eps^2 )$-fraction of all joint orders placed by the optimal policy ${\cal P}^*$. In contrast, for any frequent item $\ell$, it is generally unclear how $N_{\ell}^*$ and $N^*$ are related. To ensure an advantageous correspondence between these parameters, suppose we convert the optimal policy ${\cal P}^*$ into a revised policy ${\cal P}^{+1}$ such that, whenever a joint order is placed, it is augmented by including all frequent items. Clearly, joint ordering costs remain unchanged, i.e., $J( {\cal P}^{+1} ) = J( {\cal P}^{*} )$. In addition, the next claim shows that the marginal costs of frequent and average-dense items blow-up by a factor of at most $1 + \eps$. 

\begin{lemma} \label{lem:order_freq_anywhere}
$C_{\ell}( {\cal P}_{\ell}^{+1} ) \leq (1 + \eps) \cdot C_{\ell}( {\cal P}^*_{\ell} )$ for every item $\ell \in {\cal F} \uplus {\cal A}_{\mydense}$.    
\end{lemma}
\begin{proof}
We begin by noting that, since no modifications have been made with respect to average-dense items, $C_{\ell}( {\cal P}_{\ell}^{+1} ) = C_{\ell}( {\cal P}^*_{\ell} )$ for all $\ell \in {\cal A}_{\mydense}$. Now, focusing on some frequent item $\ell \in {\cal F}$, let us first observe that since ${\cal P}_{\ell}^{+1}$ orders item $\ell$ at all joint ordering points of ${\cal P^*}$, we necessarily have $Q( {\cal P}_{\ell}^{+1} ) \leq Q( {\cal P}^*_{\ell} )$, implying that ${\cal H}_{\ell}( {\cal P}_{\ell}^{+1} ) \leq {\cal H}_{\ell}( {\cal P}^*_{\ell} )$. Therefore, to conclude the proof, since ${\cal K}_{\ell}( {\cal P}_{\ell}^{+1} ) = K_{\ell} \cdot N( {\cal P}_{\ell}^{+1} ) = K_{\ell} N^*$, it suffices to show that $K_{\ell} N^* \leq \eps \cdot C_{\ell}( {\cal P}^*_{\ell} )$. For this purpose, by  Lemma~\ref{lem:EOQ_bound},
\begin{eqnarray}
C_{\ell}( {\cal P}_{\ell}^* ) & \geq & 2T \sqrt{ K_{\ell} H_{\ell}} \nonumber \\
& = & \frac{ 2T K_{\ell} }{ \sqrt{ \rho_{\ell} } } \label{eqn:lem_order_freq_anywhere_1} \\
& \geq & \frac{ K_{\ell} N^* }{ \eps } \ . \label{eqn:lem_order_freq_anywhere_2}
\end{eqnarray}
Here, equality~\eqref{eqn:lem_order_freq_anywhere_1} holds since $\rho_{\ell} = \frac{ K_{\ell} }{ H_{\ell} }$, whereas inequality~\eqref{eqn:lem_order_freq_anywhere_2} is obtained by recalling that $\sqrt{ \rho_{\ell} } \leq \frac{ 2\eps T }{ N^* }$, since item $\ell$ is frequent. By rearranging this bound, we indeed get $K_{\ell} N^* \leq \eps \cdot C_{\ell}( {\cal P}^*_{\ell} )$. 
\end{proof} 

\paragraph{The grids $\bs{{\cal T}_{\myzero}}$ and $\bs{{\cal T}_{\myallow}}$.} We remind the reader that, in Section~\ref{subsec:rare_grid}, our policy for rare and average-sparse items was placing orders along an evenly-spaced grid. Sticking to the high-level spirit of this construction, in order to design a replenishment policy for frequent and average-dense items, we begin by considering two grids, ${\cal T}_{\myzero} \subseteq {\cal T}_{\myallow}$, formally defined as follows:
\begin{itemize}
    \item For simplicity of notation, let $\Gamma_{\myallow} = \lceil \frac{ \eps T }{ N^* } \rceil$ and let $\Gamma_{\myzero} = \frac{ \Gamma_{\myallow} }{ \eps^4 }$, noting that the latter parameter must be an integer multiple of the former, due to initially assuming that $\frac{ 1 }{ \eps }$ takes an integer value. In addition, $\Gamma_{\myzero} < T$, as we are currently handling the difficult scenario, where $N^* \geq \frac{ 1 }{ \eps^4 }$. 

    \item The first grid ${\cal T}_{\myallow} \subseteq [T]$ consists of time points that are evenly-spaced by $\Gamma_{\myallow}$, namely, ${\cal T}_{\myallow} = \{ 1, 1 + \Gamma_{\myallow}, 1 + 2 \Gamma_{\myallow}, \ldots \}$. Clearly, $| {\cal T}_{\myallow} | \leq \frac{ N^* }{ \eps }$.
    
    \item Our second grid  ${\cal T}_{\myzero} \subseteq [T]$ is comprised of time points that are evenly-spaced by $\Gamma_{\myzero}$, namely, ${\cal T}_{\myzero} = \{ 1, 1 + \Gamma_{\myzero}, 1 + 2 \Gamma_{\myzero}, \ldots \}$. In this case, $|{\cal T}_{\myzero}| \leq \eps^3 N^*$, and moreover, ${\cal T}_{\myzero} \subseteq {\cal T}_{\myallow}$, since $\Gamma_{\myallow}$ divides $\Gamma_{\myzero}$.    
\end{itemize}    

\paragraph{Grid-aligned policies.}
Given these grids, we say that a replenishment policy for frequent and average-dense items is $({\cal T}_{\myzero}, {\cal T}_{\myallow})$-aligned when: (A)~Orders are placed only at ${\cal T}_{\myallow}$-points; and (B)~For all items, zero inventory levels are reached at every ${\cal T}_{\myzero}$-point. In what follows, we prove that this class of policies suffices to compete against the optimal policy ${\cal P}^*$ in terms of its operational cost with respect to frequent and average-dense items.

\begin{lemma} \label{lem:exists_good_aligned}
There exists a $({\cal T}_{\myzero}, {\cal T}_{\myallow})$-aligned policy ${\cal P}^{+2}$ satisfying the next two properties: 
\begin{enumerate}
    \item {\em Number of joint orders}: $N( {\cal P}^{+2} ) \leq (1 + \eps) \cdot N^*$.

    \item {\em Marginal costs}: $C_{\ell}( {\cal P}^{+2}_{\ell} ) \leq (1 + 17\eps) \cdot C_{\ell}( {\cal P}^*_{\ell} )$, for every item $\ell \in {\cal F} \uplus {\cal A}_{\mydense}$.  
\end{enumerate}
\end{lemma}
\begin{proof}
To establish the desired claim, we alter our previously-considered policy ${\cal P}^{+1}$ into a  $({\cal T}_{\myzero}, {\cal T}_{\myallow})$-aligned policy ${\cal P}^{+2}$ as follows:
\begin{enumerate}
    \item \label{item:def_P2_1} {\em Ensuring property~A}: First, for every item $\ell \in {\cal F} \uplus {\cal A}_{\mydense}$, each $\ell$-ordering point $\tau_{\ell,\nu}^{+1}$ of the policy ${\cal P}^{+1}$ is relocated to $\lceil \tau_{\ell,\nu}^{+1} \rceil^{ ({\cal T}_{\myallow}) }$, where $\lceil \cdot \rceil^{ ({\cal T}_{\myallow}) }$ is an operator that rounds its argument up to the nearest point in ${\cal T}_{\myallow}$.

    \item \label{item:def_P2_2} {\em Ensuring property~B}: Next, a joint order is placed at every point in ${\cal T}_{\myzero}$, consisting of all items in ${\cal F} \uplus {\cal A}_{\mydense}$.
\end{enumerate}

Given these modifications, in terms of joint orders, we have $N( {\cal P}^{+2} ) \leq N( {\cal P}^{+1} ) + |{\cal T}_{\myzero}| \leq (1 + \eps) \cdot N^*$, since $N( {\cal P}^{+1} ) = N( {\cal P}^* ) = N^*$ and $|{\cal T}_{\myzero}| \leq \eps^3 N^*$. Next, let us focus on some item $\ell \in {\cal F} \uplus {\cal A}_{\mydense}$ and upper-bound its  marginal operational cost $C_{\ell}( {\cal P}^{+2}_{\ell}) = K_{\ell} \cdot N( {\cal P}^{+2}_{\ell}) + H_{\ell} \cdot Q( {\cal P}^{+2}_{\ell})$. To this end, the number of $\ell$-orders placed by our policy is $N( {\cal P}^{+2}_{\ell} ) \leq N( {\cal P}^{+1}_{\ell} ) + |{\cal T}_{\myzero}| \leq (1 + \eps) \cdot N( {\cal P}^{+1}_{\ell} )$, where the latter inequality follows by recalling that $|{\cal T}_{\myzero}| \leq \eps^3 N^*$ and $N( {\cal P}^{+1}_{\ell} ) \geq \eps^2 N^*$, since $N( {\cal P}^{+1}_{\ell} ) = N^*$ when item $\ell$ is frequent, and $N( {\cal P}^{+1}_{\ell} ) = N( {\cal P}^*_{\ell} ) \geq \eps^2 N^*$ when it is average-dense. Regarding the comparison between $Q( {\cal P}^{+2}_{\ell} )$ and $Q( {\cal P}^{+1}_{\ell} )$, the important observation is that alteration~\ref{item:def_P2_1} lengthens each order duration by at most $\Gamma_{\myallow}$, which may only become shorter following alteration~\ref{item:def_P2_2}. Therefore, arguments similar to those employed for proving Lemmas~\ref{lem:grid_rare} and~\ref{lem:grid_for_sparse} show that
\begin{eqnarray}
Q( {\cal P}^{+2}_{\ell} ) & \leq &  Q( {\cal P}^{+1}_{\ell} ) + 2\Gamma_{\myallow} T + N( {\cal P}^{+1}_{\ell} ) \cdot \Gamma_{\myallow}^2 \nonumber \\
& \leq & Q( {\cal P}^{+1}_{\ell} ) + \frac{ 8\eps T^2 }{ N^* } \label{eqn:lem_exists_good_aligned_1} \\
& \leq & (1 + 8\eps) \cdot Q( {\cal P}^{+1}_{\ell} ) \ . \label{eqn:lem_exists_good_aligned_2}
\end{eqnarray}
Here, inequality~\eqref{eqn:lem_exists_good_aligned_1} holds since $\Gamma_{\myallow} = \lceil \frac{ \eps T }{ N^* } \rceil$ and $N( {\cal P}^{+1}_{\ell} ) \leq N^*$. Inequality~\eqref{eqn:lem_exists_good_aligned_2} is obtained by observing that, since ${\cal P}^{+1}_{\ell}$ places $N( {\cal P}^{+1}_{\ell} ) \leq N^*$ orders across a planning horizon of length $T$,   Lemma~\ref{lem:convex_bound} informs us that $Q( {\cal P}^{+1}_{\ell} ) \geq \frac{ T^2 }{ N( {\cal P}^{+1}_{\ell} ) } \geq \frac{ T^2 }{ N^* }$. By combining both types of costs,
\begin{eqnarray*}
C_{\ell}( {\cal P}^{+2}_{\ell}) & = & K_{\ell} \cdot N( {\cal P}^{+2}_{\ell}) + H_{\ell} \cdot Q( {\cal P}^{+2}_{\ell}) \\
& \leq & (1 + \eps) \cdot K_{\ell} \cdot  N( {\cal P}^{+1}_{\ell}) + (1 + 8\eps) \cdot H_{\ell} \cdot  Q( {\cal P}^{+1}_{\ell}) \\
& \leq & (1 + 8\eps) \cdot C_{\ell}( {\cal P}^{+1}_{\ell}) \\
& \leq & (1 + 17\eps) \cdot C_{\ell}( {\cal P}^*_{\ell}) \ ,
\end{eqnarray*}
where the last inequality holds since $C_{\ell}( {\cal P}_{\ell}^{+1} ) \leq (1 + \eps) \cdot C_{\ell}( {\cal P}^*_{\ell} )$ for every item $\ell \in {\cal F} \uplus {\cal A}_{\mydense}$, by Lemma~\ref{lem:order_freq_anywhere}.
\end{proof}

\paragraph{Computing an optimal $\bs{({\cal T}_{\myzero}, {\cal T}_{\myallow})}$-aligned policy.} In the remainder of this section, we propose an $O( 2^{ O(1/\eps^4) } \cdot (nT)^{ O(1) } )$-time  algorithm for identifying an optimal $({\cal T}_{\myzero}, {\cal T}_{\myallow})$-aligned policy ${\cal P}^{ {\cal F} \uplus {\cal A}_{\mydense}}$ with respect to the set of frequent and average-dense items. Since ${\cal P}^{+2}$ is a feasible policy in this context, based on Lemma~\ref{lem:exists_good_aligned}, we conclude that 
\[ F( {\cal P}^{ {\cal F} \uplus {\cal A}_{\mydense}} ) ~~\leq~~ F( {\cal P}^{+2} ) ~~\leq~~ (1 + \eps) \cdot K_0 N^* + (1 + 17 \eps) \cdot \sum_{\ell \in {\cal F} \uplus {\cal A}_{\mydense} } C_{\ell}( {\cal P}_{\ell}^* ) \ . \] 

To this end, let ${\cal S}_1, \ldots, {\cal S}_{|{\cal T}_{\myzero}|}$ be the sequence of time period segments stretching between successive points in ${\cal T}_{\myzero}$,  meaning that 
\[ {\cal S}_1 ~~=~~ \{ 1, \ldots, \Gamma_{\myzero} \}, \quad {\cal S}_2 ~~=~~ \{ \Gamma_{\myzero} + 1, \ldots, 2\Gamma_{\myzero} \}, \quad {\cal S}_3 ~~=~~ \{ 2\Gamma_{\myzero} + 1, \ldots, 3\Gamma_{\myzero} \}, \quad \ldots \]
Given property~B of $({\cal T}_{\myzero}, {\cal T}_{\myallow})$-aligned policies, we know that for all frequent and average-dense items, any such policy reaches zero inventory levels at every ${\cal T}_{\myzero}$-point. Recalling that ${\cal T}_{\myzero} = \{ 1, 1 + \Gamma_{\myzero}, 1 + 2 \Gamma_{\myzero}, \ldots \}$, these are exactly the left endpoints of ${\cal S}_1, \ldots, {\cal S}_{|{\cal T}_{\myzero}|}$. Therefore, in regard to identifying an optimal $({\cal T}_{\myzero}, {\cal T}_{\myallow})$-aligned policy, the only type of dependency between these segments occurs due to their cumulative number of joints orders. Consequently, by employing straightforward dynamic programming ideas, where segments are sequentially processed in arbitrary order, it suffices to resolve the next single-segment question: Given a segment ${\cal S}_m$ and an integer $N_m \leq \Gamma_{\myzero}$, determine a minimum-cost $({\cal T}_{\myzero}, {\cal T}_{\myallow})$-aligned policy across ${\cal S}_m$, subject to placing at most $N_m$ joint orders.

The important observation is that, in $O( 2^{ O(1/\eps^4) } \cdot (nT)^{ O(1) } )$ time, we can compute an optimal policy for this subproblem. Indeed, given property~A of $({\cal T}_{\myzero}, {\cal T}_{\myallow})$-aligned policies, we know that orders can be placed only at ${\cal T}_{\myallow}$-points. However, the number of ${\cal T}_{\myallow}$-points in ${\cal S}_m$ is precisely $\frac{ \Gamma_{\myzero} }{ \Gamma_{\myallow} } = \frac{ 1 }{ \eps^4 }$, meaning that we can enumerate over all $2^{ O(1/\eps^4) }$ subsets of time periods where orders could be  placed. For each such choice of cardinality at most $N_m$, we independently solve a discrete-time single-item lot sizing problem for each item $\ell \in {\cal F} \uplus {\cal A}_{\mydense}$.

\subsection{Additional proofs} \label{subsec:add_proofs_discrete}

\paragraph{Proof of Claim~\ref{clm:aux_grid_rare}: Upper bound on $\bs{\mathrm{(I)}}$.} By definition of $\psi_{\ell}$, we have $(\psi_{\ell}-1) \cdot \Gamma \leq \sqrt{ \rho_{\ell} } < \psi_{\ell} \Gamma$. In addition,  $\psi_{\ell} \geq \frac{ 1 }{ 2\eps}$, by inequality~\eqref{eqn:UB_psi_ell}, implying that $\sqrt{ \rho_{\ell} } \geq (1 - 2\eps) \cdot \psi_{\ell} \Gamma$. Based on these observations,
\begin{eqnarray}
\mathrm{(I)} & = &  \frac{ K_{\ell}T }{ \psi_{\ell}\Gamma }  + H_{\ell} T \psi_{\ell}\Gamma \nonumber \\
& < & T \cdot \left( \frac{ K_{\ell} }{ \sqrt{\rho_{\ell}}} + \frac{ 1 }{ 1 - 2\eps } \cdot H_{\ell} \sqrt{\rho_{\ell}} \right) \nonumber \\
& \leq & (1 + 4\eps) \cdot 2T \sqrt{ K_{\ell} H_{\ell}} \label{eqn:lem_grid_rare_6} \\
& \leq & (1 + 4\eps) \cdot C_{\ell}( {\cal P}_{\ell}^* ) \ . \label{eqn:lem_grid_rare_7}  
\end{eqnarray}
Here, inequality~\eqref{eqn:lem_grid_rare_6} is obtained by plugging in $\rho_{\ell} = \frac{ K_{\ell} }{ H_{\ell} }$. Inequality~\eqref{eqn:lem_grid_rare_7} follows from our EOQ-based bound in Lemma~\ref{lem:EOQ_bound}, implying in particular that $C( {\cal P}_{\ell}^* ) \geq 2T  \sqrt{K_{\ell} H_{\ell}}$.

\paragraph{Proof of Claim~\ref{clm:aux_grid_rare}: Upper bound on $\bs{\mathrm{(II)}}$.} Moving on to consider the second term, we have
\begin{eqnarray}
\mathrm{(II)} & = & K_{\ell} + H_{\ell} \cdot \left( \psi_{\ell} \Gamma \right)^2 \nonumber \\
& \leq & K_{\ell} + 4 H_{\ell} \rho_{\ell} \label{eqn:lem_grid_rare_8} \\
& = & 5 K_{\ell} \label{eqn:lem_grid_rare_9} \\
& < & 5\eps K_{\ell} N_{\ell}^* \label{eqn:lem_grid_rare_10} \\
& \leq & 5\eps \cdot C_{\ell}( {\cal P}_{\ell}^* ) \ . \nonumber
\end{eqnarray}
Here, inequality~\eqref{eqn:lem_grid_rare_8} holds since $\sqrt{ \rho_{\ell} } \geq (1 - 2\eps) \cdot \psi_{\ell} \Gamma$, as observed when upper-bounding  $\mathrm{(I)}$. Inequality~\eqref{eqn:lem_grid_rare_9} follows by recalling that $\rho_{\ell} = \frac{ K_{\ell} }{  H_{\ell} }$. Finally, inequality~\eqref{eqn:lem_grid_rare_10} is obtained by noting that $N_{\ell}^* > \frac{ 1 }{ \eps }$, according to the hypothesis of case~2.

\paragraph{Proof of Lemma~\ref{lem:grid_for_sparse}.} Suppose that the sequence of $\ell$-ordering points with respect to ${\cal P}^*_{\ell}$ is given by $1 = \tau_{\ell,1}^* < \tau_{\ell,2}^* < \cdots < \tau_{\ell,N_{\ell}^*}^*$. Then, our policy ${\cal P}^{{\cal A}_{\mysparse}}_{\ell}$ will place $\ell$-orders at the time points $1 = \lceil \tau_{\ell,1}^* \rceil^{ ({\cal T}) } \leq \lceil \tau_{\ell,2}^* \rceil^{ ({\cal T}) } \leq \cdots \leq \lceil \tau_{\ell,N_{\ell}^*}^* \rceil^{ ({\cal T}) }$. As such, in terms of ordering costs, ${\cal K}_{\ell}( {\cal P}^{{\cal A}_{\mysparse}}_{\ell} ) \leq {\cal K}_{\ell}( {\cal P}^*_{\ell} )$. As far as holding costs are concerned, similarly to case~1 in  Lemma~\ref{lem:grid_rare}, we have
\begin{eqnarray}
Q( {\cal P}^{{\cal A}_{\mysparse}}_{\ell} ) & \leq & Q( {\cal P}_{\ell}^* ) + 2\Gamma T + N_{\ell}^* \Gamma^2 \nonumber \\
& \leq & Q( {\cal P}_{\ell}^* ) + \frac{ 8T^2 }{ \eps N^* } \label{eqn:lem_grid_rare_11} \\
& \leq & Q( {\cal P}_{\ell}^* ) + \frac{ 8\eps T^2 }{ N_{\ell}^* }\label{eqn:lem_grid_rare_12}  \\
& \leq & (1 + 8\eps ) \cdot Q( {\cal P}_{\ell}^* ) \ . \label{eqn:lem_grid_rare_13}
\end{eqnarray}
Here, inequalities~\eqref{eqn:lem_grid_rare_11} and~\eqref{eqn:lem_grid_rare_12} are obtained by recalling that  $\Gamma = \lceil \frac{ T }{ \eps N^* } \rceil \leq \frac{ 2T }{ \eps N^* }$ and $N_{\ell}^* < \eps^2 N^*$, since item $\ell$ is average-sparse. Inequality~\eqref{eqn:lem_grid_rare_13} follows by noting that ${\cal P}_{\ell}^*$ places $N_{\ell}^*$ orders across a  planning horizon of length $T$, and therefore, $Q( {\cal P}_{\ell}^* ) \geq \frac{ T^2 }{ N_{\ell}^* }$ by Lemma~\ref{lem:convex_bound}.

\section{Concluding Remarks}

We believe that our work opens the door for a number of promising directions to be investigated in future research. The next few paragraphs are dedicated to highlighting some of these prospective opportunities, from those that appear quite-possibly achievable to others that pose deep theoretical challenges.

\paragraph{Running time improvements?} As expert readers may have observed while delving into the nuts-and-bolts of our analysis, to maintain accessibility and conceptual simplicity, this paper intentionally refrains from optimizing constants in $\eps$-dependent running time exponents. Specifically, it is worth emphasizing that the $2^{ \tilde{O}( 1 / \eps^4 ) }$-term we are meeting in Theorems~\ref{thm:EPTAS_dynamic_cont} and~\ref{thm:main_EPTAS} represents a reasonable compromise between performance guarantees and digestibility, rather than genuine attempts to arrive at the most efficient approximation scheme. As an appealing direction for future investigation, it would be interesting to examine whether highly refined analysis, possibly combined with additional ideas, could perhaps lead to faster implementations.

\paragraph{Approximation schemes for nearby models?} Within the domain of joint replenishment, Section~\ref{sec:reduction} devises an efficient discretization-based framework, proving that every continuous-time infinite-horizon instance can be reduced to the discrete-time finite-horizon setting, while incurring a multiplicative optimality loss of at most $1 + \eps$. We believe that the existence of analogous reductions in relation to nearby inventory management models is a very compelling direction for future research. In particular, the first candidate to examine along these lines could be the one-warehouse multi-retailer problem (see, e.g., \cite{Roundy85}, \cite{LuP94}, \cite{LeviRSS08}, and \cite{GayonMRS17}), given its structural resemblance to classical joint replenishment. Concurrently, multi-product lot-sizing problems revolving around assembly and distribution systems appear as very natural directions to pursue. By consulting the excellent survey of \cite{MuckstadtR93}, readers will discover that in the overwhelming majority of these settings, power-of-2 polices have been untouched as state-of-the-art approximation guarantees for nearly four decades.

\paragraph{Intractability results?} As outlined in Section~\ref{subsec:contributions}, we have witnessed truly outstanding progress in regard to hardness results for efficiently obtaining optimal periodic policies in discrete and continuous time \citep{SchulzT11, CohenHillelY18, TuisovY20, SchulzT22}. Regrettably, these achievements have not been replicated in the realm of optimal dynamic policies, even for presumably simpler questions, such as representing these policies in polynomial space. While it is widely believed that dynamic settings should be at least as intractable as periodic ones, lower bounds on the time/space complexity of computing optimal dynamic policies have not been devised to date, probably due to the crucial reliance on periodicity in prior work. Contributions of this form are highly sought-after for a broad range of dynamic replenishment problems, and we wish to conclude our work by emphasizing their desirability (and markedly  challenging nature) once again.

% BIB %%%%%%%%%%%%%%%%%%%%%%%%%%%%%%%%%%%%%%%%%%%
\addcontentsline{toc}{section}{Bibliography}
\bibliographystyle{plainnat}
\bibliography{BIB-JRP}

% APPENDIX %%%%%%%%%%%%%%%%%%%%%%%%%%%%%%%%%%%%%%
% \appendix
% \input{TEX-Appendix}

\end{document}